# Uncertainty spill-overs: when policy and financial realms overlap

Emanuele BACCHIOCCHI[a] and Catalin DRAGOMIRESCU-GAINA[b]

[a] University of Bologna, Department of Economics, Piazza Scaravilli 1, 40126 Bologna, Italy. Email: e.bacchiocchi@unibo.it

[b] Universita Cattolica del Sacro Cuore, Largo A. Gemelli 1, 20123 Milan, Italy. Corresponding author. Email: catalinflorinel.dragomirescugaina@unicatt.it, catalingaina@gmail.com

**Abstract**

No matter its source, financial- or policy-related, uncertainty can feed onto itself, inflicting the real economic sector, altering expectations and behaviours, and leading to identification challenges in empirical applications. The strong intertwining between policy and financial realms prevailing in Europe, and in Euro Area in particular, might complicate the problem and create amplification mechanisms difficult to pin down. To reveal the complex transmission of country-specific uncertainty shocks in a multi-country setting, and to properly account for cross-country interdependencies, we employ a global VAR specification for which we adapt an identification approach based on magnitude restrictions. Once we separate policy uncertainty from financial uncertainty shocks, we find evidence of important cross-border uncertainty spill-overs. We also uncover a new amplification mechanism for domestic uncertainty shocks, whose true nature becomes more blurred once they cross the national boundaries and spill over to other countries. With respect to ECB policy reactions, we reveal stronger but less persistent responses to financial uncertainty shocks compared to policy uncertainty shocks. This points to ECB adopting a more (passive or) accommodative stance towards the former, but a more pro-active stance towards the latter shocks, possibly as an attempt to tame policy uncertainty spill-overs and prevent the fragmentation of the Euro Area financial markets.

**Declarations of interest:** none

**Funding:** This research did not receive any specific grant from funding agencies in the public, commercial, or not-for-profit sectors.

# 1. INTRODUCTION

For a few days every January, in Davos, an exclusive alpine resort in Switzerland, global financial elite mingles with political elite, central bankers and policymakers. While this hallmark event has taken place regularly for half a century now, overlaps between financial and policy realms remain significant even if one looks further back into the history. From an analytical perspective, this leads to unexpected cross-influences that can amplify each other, especially during uncertain times. Financial stress and market uncertainties can bring changes in policies or political circumstances, as much as uncertainty stemming from policy changes creates anxieties for financial investors. No matter its source, financial or policy-related, uncertainty will feed onto itself, contaminating other areas and leading to identification challenges in empirical applications. Unfortunately, markets and investors are probably better equipped to evaluate and price risk rather than uncertainty, which is a broader concept encompassing risk and requiring different analytical approaches. We try to add to the existing stock of analytical methods able to disentangle among various sources of uncertainty but in a multi-country context, where cross-border spill-overs and overlaps are expected to pose additional identification challenges.

From this perspective, the European Union (EU), and the Euro Area (EA) in particular – with its rather incomplete institutional architecture –, make for an interesting case due to a high potential for uncertainty spill-overs and overlaps. On the one hand, domestic policy uncertainty can reverberate at the European (and global) levels with serious financial consequences measured in terms of bond yields, financial stock prices or currency moves. In June 2015 the Greek government called a snap referendum over its bailout terms, generating chaos in European policy circles, but also among financial investors who feared a Euro Area (EA) breakdown; as market sentiment turned sour, Greek sovereign bond spreads reached unprecedented levels and the country was effectively cut off global financial markets, while domestic banks suffered and were forced to impose strict capital controls. On the other hand, it is the banking sector turmoil that echoes in the policy domain, as risks are transferred from the private to the public sector due to bank-rescue packages that increase sovereign and contagion risks (see Acharya et al., 2014; Attinasi et al., 2010; Bicu and Candelon 2013; Stanga 2014). Ireland perfectly illustrates this latter case, when the government introduced guarantees to address the weakness of the domestic banking sector in September 2008, after the Lehman shock; as a result, banks' credit default swaps (CDS) came down but the Irish sovereign CDS spiked abruptly (Stanga 2014; Leonello 2018).

The present paper aims at exploring this deep and complex intertwining, which is particularly prevalent in Europe, between policy and financial realms, whose interactions might create amplification mechanisms for country-specific uncertainty shocks. Whether such mechanisms work to amplify financial uncertainty, policy uncertainty, none, or both is the most important research question we address in this paper. We seek therefore to contribute to a new and rapidly expanding literature strand that deals with various uncertainty measures, their sources, effects, and cross-border spill-overs (see among many others Bekaert et al., 2013; Caldara et al., 2016; Bacchiocchi 2017; Shin and Zhong 2018; Ludvigson et al., 2019; Angelini et al., 2019).

We also aim at understanding what specific role the European Central Bank (ECB) has played in counteracting various uncertainty sources and/or their spill-overs at the European level. Over the last decade, the ECB considerably expanded its policy toolkit, took greater supervisory and regulatory duties, and stepped in when there was no credible policy actor for global financial markets[1], up to the point of being called 'the only game in town'.[2] On the back of a rather complicated EA governance structure, the ECB provided an effective backstop to area-wide financial stress, while treating country-specific shocks with more flexibility. This is in spite of the fact that, in many instances, country-specific factors have penetrated the decision-making process in both Brussels and Frankfurt.

Complex identification challenges arise within a multi-country setting, such as the EA, due to its significant financial integration, but incomplete political integration, and where information frictions play an important role (see Freixas and Holthausen, 2004). During the European sovereign debt crisis, domestic banks in some EA periphery were given incentives to draw more central bank liquidity, largely against domestic sovereign bonds. Battistini et al., (2014) and Acharya and Steffen (2015) provide empirical evidence on these mechanisms, where bailed-out periphery banks hold more periphery sovereign debt.[3] Recently, the theoretical work of Farhi and Tirole (2017), Leonello (2018), and Cooper and Nikolov (2018) shed light on the feedback-loops between sovereigns and banks, but strong feedback-loops can blur the thin separation line between financial and policy realms.

Understanding the ECB role is an important topic because the current EA governance system suffers from a lack of institutional leadership to deal with several uncertainty sources. The existing literature on (monetary and fiscal) policy interactions within a common currency area does not provide us with sufficient clarifications in this regard (for a recent survey, see Foresti, 2018). ECB faces numerous and delicate policy trade-offs in pursuing its price stability mandate, set according to the EU Treaties. A clearer distinction between policy and financial uncertainty shocks could improve ECB policy effectiveness, and even shield it from possible legal actions.[4] There have been many controversies surrounding ECB monetary policy conduct, especially with respect to its unconventional measures, like the various asset purchasing programs implemented over the last decade. In August 2011, for example, the Securities Markets Programme (SMP) made some sizeable bond purchases from the EA periphery, especially Italian and Spanish sovereigns, with some positive effects on spreads in unsettled market conditions. However, the program was soon suspended for Italian bonds as it became clear that the Berlusconi government was not delivering on its promised economic reforms; fast forward in November 2011, market confidence in the Italian government collapsed and a new prime-minister was appointed.

---

[1] Mario Draghi's speech on 26th July 2012 has been considered a cornerstone moment for the EA sovereign debt crisis. See https://www.ecb.europa.eu/press/key/date/2012/html/sp120726.en.html.
[2] See Otmar Issing's comment at: https://www.centralbanking.com/central-banks/economics/2473842/otmar-issing-on-why-the-euro-house-of-cards-is-set-to-collapse.
[3] There are plenty of other empirically relevant studies on the moral hazard prevalent during the European sovereign debt crisis. Acharya et al., (2014) show that CDS for sovereigns and banks commove over the European crisis period, but not much before the crisis. Koijen et al., (2017) document the home bias existing in vulnerable countries during the implementation of the ECB asset purchasing programmes.
[4] See the decision of the Court of Justice of the European Union in favour of the ECB's Public-Sector Purchase Programme (PSPP) at https://www.reuters.com/article/us-ecb-policy-court/ecb-wins-courts-backing-for-buying-government-debt-idUSKBN1OA0Q0. This decision stands in contrast to a more recent decision of the German Constitutional Court, thus raising unprecedented legal challenges for the EU governance system.

Given these large consequences stemming from the interaction of financial and policy realms in the EA, it is important to know whether there are sizable spill-overs of country-specific uncertainty shocks, and whether ECB can play any specific role. Our main contribution is to approach these important research questions from an empirical perspective that is able to deal with the inherent identification challenges that arise in a multi-country setting. In this respect, we adopt and extend the identification approach based on magnitude restrictions, recently proposed in De Santis and Zimic (2018), to a global vector autoregressive (henceforth GVAR) setting. There are other few distinct but comparable approaches in a rapidly expanding empirical literature aiming at identifying (different types of) uncertainty shocks (e.g. Bacchiocchi 2017; Shin and Zhong 2018; Ludvigson et al., 2019; Angelini et al., 2019); as each methodological approach has its own merits, we regard them as largely complementary to ours. Inspired by event studies, the identification based on magnitude restrictions was proposed by De Santis and Zimic (2018) to expose spill-overs between U.S. and European sovereign bond yields. However, it is quite general and allows for the identification of shocks from within any strongly correlated set of variables, like for example in cases of statistical overlaps. An important aspect for its application in our case is that the two uncertainty proxies focus on distinct data sources, and rely on different measurement approaches.

To capture financial uncertainty, we use the Composite Indicator for Systemic Stress (CISS), a highly relevant policy indicator for ECB, which also makes this indicator available on a weekly frequency, and for most EU Member States (see Hollo et al., 2012). Compared to other financial uncertainty measures that are probably more readily available (e.g. CDS, volatility, cross-sectional variation), composite indicators summarize a higher dimensional space and are more efficient in reflecting financial stress across several market segments.[5] Broader (or Knightian) uncertainty, instead, stemming from changes in the political landscape, rhetoric, opinions and policies is harder to measure (see discussion in Bekaert et al., 2013; Jurado et al., 2015; Baker et al., 2016; Ludvigson et al., 2019). The recent literature is booming with different measures for this type of uncertainty, spanning different methodologies and data sources. However, some of the best known indicators rely heavily on media sources. In a highly influential paper, Baker et al., (2016) propose an *economic policy uncertainty* (EPU) measure based on the frequency of some relevant keywords in major newspapers (and other commonly available media sources); they further show their indicator is orthogonal to other common measures of risk and uncertainty, such as forecasts dispersion or financial volatility etc. Because of its wide availability for different EU and EA countries, and its robustness in empirical applications, we rely on EPU as a measure of (economic) policy uncertainty.[6] Despite possible overlaps that might be revealed by a high correlation during specific periods of time, CISS and EPU draw on different data sources, measurement and index construction methodologies.[7]

---

[5] Various studies, such as Fratzscher et al., (2016), Moder (2017), Burriel and Galesi (2018), Boeckx et al., (2017) use the CISS index proposed in Hollo et al., (2012) to uncover transmission channels and consequences of financial stress across European markets.

We contribute to this growing literature strand by investigating the dynamics of uncertainty arising from the interaction of financial and policy realms, where EA stands, unfortunately, as a fertile ground for investigations. In this context, we uncover an important amplification mechanisms for domestic uncertainty shocks: once they cross the national boundaries and spill over to other EA countries, their true nature and origin can become blurred, leading to greater overlaps. This transmission mechanism, with a strong international dimension, appears novel to the existing literature on uncertainty-related topics. However, some of its theoretical ingredients can be found in studies dealing with financial integration (e.g. Freixas and Holthausen, 2004) and financial markets contagion (King and Wadhwani, 1990; Kodres and Pritsker, 2002) that feature strong cross-border information asymmetries.

The remaining of the paper is organised as follows. Section 2 discusses the theoretical background relevant for our empirical analysis. Section 3 presents the data and the modelling approach. Section 4 provides a detailed overview of the main results and their policy implications. Finally, section 5 concludes. More detailed results from our analysis are presented in the Appendixes.

## 2. THEORETICAL BACKGROUND

This section discusses the two main literature strands that most closely relate to our empirical model. Firstly, we discuss the sovereign-bank nexus, and secondly, the role of cross-border information frictions (or asymmetries) as uncertainty sources. The sovereign-bank nexus is important because it can capture the most relevant interactions between financial and policy realms in a single-country setting. However, in multi-country settings, these theories might not adequately explain the multiplicity of the mechanisms and interactions coexisting across EA.

---

[6] Among others, Benati (2014) estimates the impact of EPU shocks in the Euro Area, US, UK and Canada over a sample that includes the Global Financial Crisis; Ludvigson et al., (2019) employ EPU along with other measures of uncertainty in order to identify the consequences of uncertainty shocks on real output; see also Stock and Watson (2012), Caldara et al., (2016), Choi and Furceri (2019) who use EPU for robustness checks.

[7] A related literature strand employs sovereign and banking risk measures derived directly from market dynamics and prices, like from CDSs, where identification challenges are small and identification is straightforward (see Bicu and Candelon 2013; Stanga 2014; Acharya et al., 2014; Greenwood-Nimmo et al., 2019; Bettendorf 2019).

**2.1. Sovereign-bank nexus**

The sovereign-bank nexus, which is defined as the interaction between the financial and policy realms, is one of the main uncertainty sources in financial economics. What we are most interested in learning about is the very first stage of the uncertainty generating process, where policy and financial uncertainty usually combine and amplify each other, leading to identification challenges in empirical works. Once uncertainty arises, it propagates rapidly and inflicts the real sector affecting investment dynamics, asset prices, firms' balance sheets, credit spreads etc., amplified mainly by financial frictions (see among many others, Arellano et al., 2010; Christiano et al., 2014; Bloom 2014; Gilchrist et al., 2014; Bloom et al., 2018).[8]

The main theoretical mechanisms underpinning the feedback loops between banks and sovereigns are best described in Farhi and Tirole (2017), Faia (2017), Leonello (2018), Allen et al., (2018), Cooper and Nikolov (2018). For the sake of the discussion, we only briefly summarize the two key mechanisms featuring in these models. On the one hand, since banks hold sovereign bonds in their books for liquidity and regulatory reasons, sovereign distress can contaminate the banking sector. On the other hand, the (implicit or explicit) guarantees provided by the government allow banking sector distress to inflict the public sector. Empirical evidence on these theoretical transmission mechanisms is provided, among many others, in Bicu and Candelon (2013), Stanga (2014), Bettendorf (2019). While the evidence is clear, in reality there are some nuances one needs to consider. Government commitment to bailing out the banking sector depends on its fiscal capacity and debt dynamics, but these constraints can fail due to moral suasion, which explains why EA periphery banks had higher levels of domestic sovereign bonds in their books (Acharya et al., 2014; Koijen et al., 2017; Greenwood-Nimmo et al., 2019). Besides the fiscal costs of a bailout, the central bank can be involved along with the government, in which case there will be inflation and devaluation costs (Farhi and Tirole, 2017).

These theoretical models describing the sovereign-bank feedback loops are all set within a single-country framework, and therefore cannot be easily extended to a multi-country setting, which is the main focus in this paper. Difficulties arise from the lack of full political integration across the EA, and in particular the lack of a fully-fledged Banking Union. Recent institutional reforms at the EU level are welcome, although a lack of political consensus is hindering further progress in this direction.[9]

---

[8] Empirical evidence on these transmission mechanisms is provided in Stock and Watson (2012); Caldara et al., (2016), Alessandri and Mumtaz (2019).
[9] A Single Resolution Mechanism working in conjunction with a Single Supervisory Mechanism (SSM) were recently established (second half of 2014), under the coordination of the ECB, together with competent supervisory authorities from EA Member States. These are two of the three pillars required for the Banking Union to function effectively; the third pillar, i.e. a common deposit guarantee across the entire EA, is still missing. For more information, see: https://ec.europa.eu/info/business-economy-euro/banking-and-finance/banking-union_en.

## 2.2. Cross-border information frictions

For a multi-country perspective, a change in focus is in order. Without a fully operational Banking Union and a complete political integration, the theoretical mechanisms underlying the sovereign-bank nexus do not directly apply at the EA level. Therefore, focusing on domestic bond holdings and government guarantees is no longer sufficient; instead, the focus must fall on the rebalancing of international portfolios across financially-integrated markets and the main factors driving this reallocation process, such as cross-border information frictions and asymmetries.

European cross-border banking has dramatically increased financial integration as a direct result of the two banking directives adopted in 1977 and 1989 that aimed at eliminating restrictions, harmonizing regulation, and achieving better coordination in prudential supervision. Besides its benefits measured in terms of reduced costs and access to financial services, it was hoped that integration would have increased the effectiveness of ECB monetary policy and improved its transmission mechanisms.[10] Next, we will review theories suggesting that financial integration does not necessarily reduce information frictions, but might even increase financial fragility.

Freixas and Holthausen (2004) show that integration of the EA interbank market can magnify the asymmetry of information in cross-border banking, creating a contagion channel and financial fragility. Depending on the amount of information frictions, their model allows for multiple equilibria. In particular, the model differentiates between financial *segmentation* and *integration*, where the former relates to a case where all interbank transactions occur within the national borders, liquidity distribution is inefficient and interest rates are higher, while the latter refers to the opposite case. The main theoretical insights from Freixas and Holthausen (2004) are that a segmented market equilibrium is always possible, but an integrated market equilibrium is not necessarily feasible at all times; sometimes, they find that the integrated market equilibrium is not even welfare improving due to increased financial fragility. In fact, more recently, Passari and Rey (2015) conclude that large welfare gains from financial integration, in general, are rather hard to find (in contrast to the earlier findings from Allen et al., 2011). According to Freixas and Holthausen (2004), asymmetries leading to market segmentation arise when information remains locally bounded, like in the case of substantial differences in cultures and accounting practices (e.g. policy decisions to restrict risk modelling options for banks), or in local policy preferences with respect to prudential supervision (e.g. commitment to bail out a bank in distress). These few examples point to uncertainty sources that originate in the policy realm.

More recently, Garleanu et al., (2015) present a theoretical model where access to (foreign and domestic) financial markets is subject to information frictions, which lead to limited market integration in equilibrium. Moreover, because portfolio diversification (i.e. participation in distant markets) and leverage are complements in their model, a symmetric equilibrium might fail to exist, just as in Freixas and Holthausen (2004). In reality, the potential for diversification benefits within the EA is a complex function of the delicate balance between common and idiosyncratic dynamics affecting international portfolios. For example, holding EA periphery versus EA core bonds brought substantial profits for European banks, an investment strategy that Acharya and Steffen (2015) have labelled as "the 'greatest' carry trade ever". These situations point to the financial realm as a potential source of uncertainty, with information frictions and asymmetries playing an amplifying role.

---

[10] Legislative proposals to advance the integration of European capital markets, along with other segments of the financial market, are high on the policy agenda in Brussels and Frankfurt. Overall, financial integration had positive welfare effects over the first decade of the common currency, as summarized in Allen et al., (2011).

A closely related literature strand focuses on the role of cross-border information asymmetries in models of financial contagion, like for example in King and Wadhwani (1990) and Kodres and Pritsker (2002). If two countries share a common macroeconomic factor, then portfolio rebalancing by informed investors in response to a domestic shock in one country will show up as a change in demand for assets in the other country. However, as long as some investors remain uninformed and cannot correctly infer the source of the latter change in asset demand (i.e. either due to rebalancing, or due to local idiosyncratic news), cross-border information asymmetries would work by amplifying market volatility (and ultimately lead to contagion). For our setting, it is important to note that investors' confusion with respect to the nature and origin of the shock triggering excess market moves rests on the presence of strong cross-border information asymmetries.

In summary, while each of these theoretical mechanisms has its own potential merits, there is not a single one that can explain the complex, dynamic, double causality influences arising between financial and policy uncertainty within the EA. Starting from this reasoning, our empirical exercise can be seen as an attempt to shed light on the strength of these interactions that can have important consequences and policy implications.

## 3. DATA AND METHODOLOGY

### 3.1 The GVAR model

The global vector autoregressive model, or GVAR, was designed to simultaneously model cross-sectional dependence and time-series behaviour in macroeconomic data. This very flexible empirical framework was originally proposed by Pesaran et al. (2004), and extended by Dees et al. (2007). In essence, the GVAR is a collection of country-specific vector autoregressive models (or VARs), conveniently linked via a weighting matrix that makes the estimation feasible by reducing the parameter space. As we will discuss later in the section, we use financial weights based on IMF CPIS data, which reflect the importance of financial flows in explaining the dynamics of sovereign bond yield spreads, and the transmission of uncertainty spill-overs.

In principle, the GVAR model embeds three channels of cross-country interactions through: (i) foreign-specific variables, (ii) common factors and (iii) contemporaneous dependence of shocks. In this section, we allow for foreign-specific (or so-called star, i.e. *) variables to interact with domestic ones via the first channel, while in the next section, we introduce the second channel that works through common variables (i.e. the ECB monetary policy proxies). The third channel is implicitly accounted for through the estimated variance-covariance matrix in both this section and the next one. As long as the pairwise cross-country correlations left in the model residuals are low, most GVARs in the literature capture the cross-country interactions only through the first two channels, restricting[11] the variance-covariance matrix to be block-diagonal (e.g. Cesa-Bianchi, 2013; Eickmeier and Ng, 2015; Feldkircher and Huber, 2016). However, since our focus is specifically on uncertainty spill-overs, we would like to capture the second-order moments of the data as well, and therefore leave the variance-covariance matrix unrestricted in the following analysis.

In the basic GVAR specification, each country $i$ is represented by a country-specific VAR model denoted as VARX $(p_i, q_i)$, with $p_i$ and $q_i$ lags, and $Y_{i,t}$ a vector of endogenous variables. Each country-specific model is specified as:

$$Y_{i,t} = a_i + \sum_{j=1}^{p_i} B_{i,j} Y_{i,t-j} + \sum_{j=0}^{q_i} C_{i,j} Y^*_{i,t-j} + v_{i,t} \qquad (1)$$

where $a_i$ is a vector of intercepts; $B_{i,j}$ and $C_{i,j}$ are coefficient matrixes; and $v_{i,t}$ is a vector of idiosyncratic shocks, serially uncorrelated and with full variance-covariance matrix. The vector of endogenous variables $Y_{i,t}$ includes domestic variables, while foreign variables are denoted by $Y^*_{i,t} = \sum_{i \neq h} w_{i,h} Y_{h,t}$, which are specific to each country $i$ and are constructed as weighted averages of country-specific endogenous variables using a matrix of weights, $W$, where for each $i$ we have $\sum_{i \neq h} w_{i,h} = 1$.

In order to solve the GVAR, we can exploit the fact that foreign variables are linear combinations of the complete set of domestic variables $Y_t$, , i.e. $Y^*_{i,t} = W_i Y_t$, being $W_i$ the appropriate country-specific link matrix based, as we will see below, on CPIS portfolio weights. Starting from eq. (1), if we define $G_{i,0} = [I, -C_{i,0}]$ and $G_{i,j} = [B_{i,j}, C_{i,j}]$, for $j = 1, \dots, p = max(p_i, q_i)$, for each country $i$ we can obtain the alternative notation:

$$G_{i,0} W_i Y_t = a_i + \sum_{j=1}^{p} G_{i,j} W_i Y_{t-j} + v_{i,t}.$$

By staking all countries together, and denoting by $p = \max_i(\max(p_i, q_i))$ we obtain:

---

[11] In technical terms, this assumption would amount to a lack of contemporaneous volatility spill-overs between the countries included in the sample, though it would still allow for indirect volatility spill-overs that work through the complex lag structure of the model.

$$G_0 Y_t = g_0 + \sum_{j=1}^{p} G_j Y_{t-j} + v_t \qquad (2)$$

where $G_0 = \begin{pmatrix} G_{1,0} W_1 \\ \vdots \\ G_{N,0} W_N \end{pmatrix}$, $G_j = \begin{pmatrix} G_{1,j} W_1 \\ \vdots \\ G_{N,j} W_N \end{pmatrix}$, $g_0 = \begin{pmatrix} a_1 \\ \vdots \\ a_N \end{pmatrix}$ and $v_t = \begin{pmatrix} v_{1,t} \\ \vdots \\ v_{N,t} \end{pmatrix}$, with $N$ representing the number of countries. Provided that $G_0$ is invertible, we can write the GVAR in its reduced form as:

$$Y_t = h_0 + \sum_{j=1}^{p} H_j Y_{t-j} + u_t \qquad (3)$$

where $h_0 = G_0^{-1} g_0$, $H_j = G_0^{-1} G_j$ are coefficients, and $u_t = G_0^{-1} v_t$ are reduced form residuals with unrestricted covariance matrix given by $\Omega_u$. This specification of the model allows to understand the dynamic properties of the data, as well as the response of each variable in each country to a particular shock. The next step, discussed in Section 3.3, is the strategy used for the identification of the structural shocks, starting from the obtained residuals $u_t$, and, specifically, from the information contained in the covariance matrix $\Omega_u$.

### 3.2 Data and specification of the model

Our dataset focuses on the European region that is represented by 24 individual countries and one aggregate, to which we add U.S., as summarised in Table 1 below. Given the limitations of our dataset, the EA region comprises 14 individual Member States and one aggregate, i.e. the Baltics.[12] Similarly, the EU includes 20 individual countries, and the Baltics group.[13] Outside EU, we consider Russia, Turkey, Norway and Switzerland due to their strategic importance (e.g. for economic, financial, geopolitical reasons). Finally, we include U.S. as a global financial centre and an important source of macroeconomic dynamics relevant for Europe, and EA.

---

[12] Slovenia and Slovakia joined EA in 2007 and 2009 respectively, therefore, very early in the sample and before the European sovereign debt crisis. The Baltics joined the EA between 2011 and 2015, but we consider them part of the EA given their small relative size, highly open economies, and their participation in the European Exchange Rate Mechanism (ERM II) since mid-2000s – underlining the importance of ECB monetary policy for their economies.

[13] Romania, Bulgaria and Croatia suffer from limitations on data availability and were not included in the analysis; aggregating these countries is not feasible due to their larger heterogeneity than in the case of Baltics.

**Table 1:** Countries included in the empirical analysis

| Euro Area, EA | Other EU, non-EA | Others |
|---|---|---|
| Austria, AT | Czech Republic, CZ | United States, US |
| Belgium, BE | Hungary, HU | |
| Finland, FI | Poland, PL | |
| France, FR | Sweden, SE | |
| Germany, DE | Denmark, DK | |
| Italy, IT | United Kingdom, UK | |
| Ireland, IE | | |
| the Netherlands, NL | **Non-EU Europe** | |
| Spain, ES | Norway, NO | |
| Greece, EL | Switzerland, CH | |
| Portugal, PT | Turkey, TR | |
| Luxemburg, LU | Russia, RU | |
| Slovakia, SK | | |
| Slovenia, SI | | |
| Baltics, BA | | |

Note: Due to data limitations for specific indicators, we aggregate Latvia, Lithuania and Estonia into a single group, denoted as "Baltics". All indicators pertaining to Baltics are simple averages of available indicators.

Our sample consists in monthly time-series running from January 2003 to June 2018 (all data description and definitions are provided in Appendix 1). Although CISS is available with a weekly frequency from the ECB data warehouse, EPU indicators are available only with a monthly frequency. We believe that such a frequency is sufficient to uncover the most relevant spill-overs and cross-influences between the financial and policy uncertainty, due to the latter rather complex concept and measurement methodology. All country-specific EPU indexes have been calculated based on the same approach detailed in Baker et al., (2016), who propose searching major media outlets in order to gauge the frequency of some relevant keywords pertaining to the *economic policy uncertainty* domain. Obviously, speculations about un-announced policy changes, intentions or political declarations can be read almost daily in some economic and business publications, but time is of essence in order to observe sufficient political tensions that eventually feature prominently in the news (and get captured in the EPU). Considering our sample, EPU time-series[14] are available for the following 11 countries: FR, DE, NL, ES, IT, EL, IE, SE, UK, RU and US. More importantly, both EPU and CISS are available in some countries that have taken the centre stage in various EU policy debates over the last two decades (e.g. EL, IT, ES, FR, IE, UK)[15], allowing us to focus our identification strategy in the next section.

---

[14] We download all EPU time-series from www.policyuncertainty.com.
[15] Notice that only Portugal is missing from the list of so-called GIIPS (or PIIGS) countries.

Besides the EPU and CISS uncertainty proxies, the model includes the spread in 10-year sovereign bond yields against Germany, which is the analytical benchmark for the EA.[16] Including bond yield spreads along with uncertainty proxies captures the inherent trade-off between risks and returns. Working with bond yield spreads against Germany should wipe out a large part of EA-aggregate uncertainty, ensuring we capture country-specific dynamics. The rich GVAR specification allows us to further reduce the danger that our results are driven by aggregate dynamics. This happens through the inclusion of foreign variables $Y_{i,t}^*$ and common global factors, such as the VIX index[17] – which is a proxy for global risk appetite in the literature on global financial cycles (see Rey, 2015; Bruno and Shin, 2014; Miranda-Agrippino and Rey, 2015) as well as in the literature on global financial spill-overs (Chudik and Fratzscher, 2011; Bettendorf, 2019).

The original idea behind the GVAR specification is the complex re-weighting of country-specific VARs that reduces the parameters space and makes its estimation feasible (see Pesaran et al., 2004; Dees et al., 2007). To this end, we use a weighting scheme derived from data on bilateral portfolio exposures taken from the IMF's Coordinated Portfolio Investment Survey (CPIS), which includes cross-border investments in bonds and equities.[18] Due to some data limitations and mainly to streamline the interpretation of results, we use a fixed rather than a time-varying weighting matrix,[19] although the latter would probably only amplify the effects we uncover, particularly because higher capital outflows are associated with spikes in contagion risk (and uncertainty spill-overs). Compared to capital flows driven by changes in cross-border banking exposures (which we use as a robustness check), portfolio flows are less volatile, increasing the empirical relevance of our weighting scheme.[20]

Capital flows in general echo the risk-return trade-offs arising across limitedly-integrated markets (see discussion in Garleanu et al., 2015). Therefore, our modelling approach reflects the link between international capital flows and changes in sovereign spreads through the international portfolios rebalancing channel. According to this literature strand (see Rey, 2015; Bruno and Shin, 2014; Cerutti et al., 2017; Choi and Furceri, 2019), global capital flows co-move with global risk factors and monetary policy changes in centre countries like U.S. and EA. By amplifying the effects of foreign shocks on the domestic economy, capital flows can limit the policy options available to governments (Dragomirescu-Gaina and Philippas, 2015) and/or financial supervisory authorities (Allen et al., 2011), further increasing policy uncertainty. The main ingredients of these mechanisms are mirrored in our empirical specification, which includes aggregate uncertainty (i.e. weighted averages of EPU and CISS), global risk proxies (i.e. VIX), yield spreads, (weights based on) capital flows, and ECB policy proxies.

---

[16] See Appendix 1 for more data details. As a robustness check, we rebase all spreads against U.S., which represents, instead, the global benchmark (see section 4.5).
[17] VIX is the implied volatility of the S&P500 stock index and it based on option prices (source: Chicago Board of Options Exchange).
[18] Data source is http://data.imf.org/cpis. We average annual data over the 2000-2015 period (subject to availability; some countries, e.g. Baltics, had shorter time-series). The matrix is illustrated in Appendix 2.
[19] Large part of the GVAR literature simply employs fixed rather than time-varying weighting matrixes focusing more on the interactions of the model variables, rather than on weights.
[20] A similar weighting scheme based on CPIS data is employed, for example, in Hebous and Zimmermann (2013) and Greenwood-Nimmo et al. (2019). Most GVARs instead use weighting schemes based on bilateral trade flows. Eickmeier and Ng (2015) investigate several weighting schemes (e.g. based on bilateral trade, portfolio investment, foreign direct investment, banking exposures) and find that a combination between trade and financial weights works best to expose credit supply shocks in a GVAR model including real and financial variables. See also Feldkircher and Huber (2016) for an analysis of different weighting schemes in GVARs.

To account for ECB monetary policies, we can extend the specification given in equation (1) such that ECB can be considered a synthetic country in the GVAR[21] (see also Georgiadis, 2015; Burriel and Galesi, 2018). To highlight the ECB role in the model, we can re-specify the GVAR equation (1) as:

$$Y_{i,t} = a_i + \sum_{j=1}^{pi} B_{i,j} Y_{i,t-j} + \sum_{j=0}^{qi} C_{i,j} Y^*_{i,t-j} + \sum_{j=0}^{qi} D_{i,j} X_{t-j} + v_{i,t} \qquad (4)$$

where $Y_{i,t}$ and $Y^*_{i,t}$ are country-specific endogenous variables as before, but $X_t$ denotes the common variables summarising the ECB monetary policy, while $D_{i,j}$ are the associated coefficient matrices. Accordingly, the VARX associated with ECB will specify $X_t$ as an autoregressive process with lag orders given by $(px, qx)$ as:

$$X_t = m_x + \sum_{j=1}^{px} N_j X_{t-j} + \sum_{j=0}^{qx} P_j \tilde{Y}_{t-j} + v_{x,t} \qquad (5)$$

where $m_x$, $N_j$ and $M_j$ are (matrix) coefficients, $v_{x,t}$ is a noise term, and $\tilde{Y}_t$ is a vector of feedbacks from the GVAR main endogenous variables, similar to Georgiadis (2015), and Burriel and Galesi (2018). Equation (5) balances persistence with feedback effects, meaning that it can be seen as a type of reaction function where ECB responds to developments in the EA region with respect to sovereign spreads, as well as uncertainty dynamics.[22]

We follow Boeckx et al. (2017) and Burriel and Galesi (2018) and define $X_t$ above such as to capture the main aspects of the ECB policy toolbox. More specifically, we include: (i) a proxy for conventional monetary policy, denoted as $CMP$, (ii) a liquidity proxy, denoted as $Liquidity$, and (iii) an unconventional monetary policy proxy, denoted as $UMP$ (see data description in Appendix 1). In particular, we proxy $CMP$ using the Main Refinancing Operations (MRO) interest rate, which is the ECB main policy rate. As a liquidity proxy we use the spread between EONIA (i.e. the Euro Overnight Index Average) and the MRO rate. As $UMP$ proxy we use the annual change in the (log of) ECB balance sheet, which has become the standard indicator in the literature on unconventional monetary policy.

---

[21] In technical terms, equation (5) describes the dynamics of the dominant unit of the GVAR.
[22] Goldstein et al., (2011) highlight the key role of uncertainty in driving policy responses of a central bank that has imperfect information about the economic fundamentals, but can learn from market data. In our setting, even if ECB has significantly extended its regulatory and prudential oversight, the heterogeneous dynamics and fragmentation of the EA financial system implies substantial information gains if uncertainty is reduced.

Regarding the specification of the other country-specific VARs, the domestic $Y_{i,t}$ vector includes three variables for all EU countries, i.e. EPU, CISS, and the 10-year sovereign yield spread against Germany, denoted as *spread*. Obviously, three is the maximum cross-sectional dimension of the $Y_{i,t}$ vector for EA and EU countries; this happens because for some countries there is no EPU index readily available and, for Germany the sovereign spread is exactly zero, and so it is excluded as a variable. For non-EU countries, the vector $Y_{i,t}$ of domestic variables does not include CISS because ECB does not compute a CISS index for these countries. For US, instead, the vector of endogenous variables includes: the *spread*, EPU and VIX, which serves as a global proxy for risk.

The foreign vector $Y_{i,t}^*$ includes the foreign counterparts of domestic variables, along with VIX and ECB monetary policy proxies, which feature only in the case of EA countries. There are some additional particularities that we need to address, though, since not all countries belong to the same group. We choose a richer specification for EA and EU countries and impose some inherent symmetry (in terms of treating the two uncertainty proxies) in order to reflect the common European policy-making framework (i.e. through $EPU^*$), and the common financial regulatory framework (i.e. $CISS^*$). Except for US, where it is endogenous, VIX features in the foreign vector $Y_{i,t}^*$ of all other countries. Given its dominant global position, there is no $Y_{i,t}^*$ specified for the US model. The specification of $Y_{i,t}^*$ is intentionally kept simple for all other non-EU countries, such that it helps in filtering out all non-EU specific sources of noise from the model dataset.

In summary, omitting the time subscript, each VARX of the GVAR is specified as follows:

$$\text{ECB: } X = \begin{bmatrix} CMP \\ Liquidity \\ UMP \end{bmatrix} \text{ and } \tilde{Y} = \begin{bmatrix} EPU^* \\ CISS^* \\ spread^* \end{bmatrix}$$

$$\text{EA countries}^{23}: Y_i = \begin{bmatrix} EPU \\ CISS \\ spread \end{bmatrix} \text{ and } Y_i^* = \begin{bmatrix} EPU^* \\ CISS^* \\ spread^* \\ X \\ VIX \end{bmatrix}$$

$$\text{Other EU, non-EA countries: } Y_i = \begin{bmatrix} EPU \\ CISS \\ spread \end{bmatrix} \text{ and } Y_i^* = \begin{bmatrix} EPU^* \\ CISS^* \\ spread^* \\ VIX \end{bmatrix} \quad (6)$$

$$\text{Non-EU countries, other than US: } Y_i = \begin{bmatrix} EPU \\ spread \end{bmatrix} \text{ and } Y_i^* = \begin{bmatrix} spread^* \\ VIX \end{bmatrix}$$

$$\text{US: } Y_i = \begin{bmatrix} EPU \\ spread \\ VIX \end{bmatrix}$$

---

[23] The only exception here is Germany, for which: $Y_i = \begin{bmatrix} EPU \\ CISS \end{bmatrix}$, but $Y^*{}_i$ is similar as for other EA countries.

Based on these particular specifications of country-specific VARs, and following Pesaran et al. (2004), we estimate the parameters of the reduced-form GVAR in eq. (3), with the inclusion of ECB as a synthetic country as in system (6), and use the information contained in the covariance matrix $\Omega_u$ to identify the structural shocks as described in the following section.

### 3.3 Identification of structural shocks through magnitude restrictions

As noted in Dees et al (2014), and also in Dungey and Osborn (2014), dealing with multi-country models, in general, requires a different framework for conceptualizing the nature of shocks that one wishes to identify, particularly because of the strong cross-sectional dimension of these models. This is one of the contributions we bring to the uncertainty-related empirical literature, which deals largely with shock identification in single country models (noteworthy exceptions are Bicu and Candelon, 2013; Stanga, 2014; Acharya, et al., 2014; Bacchiocchi, 2017; Greenwood-Nimmo, et al., 2019; Bettendorf, 2019). A GVAR specification can elegantly solve such challenges through the inclusion of country-specific foreign variables (and common variables) that effectively reduce the cross-sectional correlation of residuals. In particular, the largest cross-sectional correlation in our GVAR is 0.19, and the corresponding average is 0.05 (in absolute terms).[24] Such small correlations, however, cannot be completely neglected in a study dealing with spill-overs.

In terms of identification, we extend De Santis and Zimic (2018) and implement structural identification through absolute magnitude restrictions in a multi-country framework. Any structural identification requires a mapping from reduced-form shocks, $u_t$, into structural ones, $\varepsilon_t$, say in the form: $u_t = S\varepsilon_t$, where $S$ is a matrix that is the focus of any identification strategy. If we normalize the structural shocks to have unit variance $E(\varepsilon_t \varepsilon_t') = I$, then we have that $\Omega_u = SS'$. A candidate for $S$ can be obtained by orthogonalizing the reduced form residuals through a rotation of the Cholesky factor of $\Omega_u$ as in Uhlig (2005) or Bacchiocchi and Kitagawa (2020), i.e. $S = \Omega_{tr}Q$, for an orthogonal matrix $Q$. Focusing on this latter, unfortunately the rotation matrix $Q$ is not unique, unless further (e.g. zero or sign) restrictions are imposed.

In practice, in our model, for each country, we are only interested in the identification of the two uncertainty shocks associated with the two uncertainty proxies. The identifying constraints we use are in the form of magnitude restrictions as in De Santis and Zimic (2018). Such restrictions work by conveniently constraining the space where the columns of interest of $S$ are required to lie. Restricting the analysis on a single country, or on a standard VAR, as in De Santis and Zimic (2018), we assume that the relative size of the contemporaneous response of uncertainty variable $i$ to an uncertainty shock $j$, with $i \neq j$, must be smaller (in absolute terms)[25] than the contemporaneous response of uncertainty variable $j$ to the same uncertainty shock $j$. In other words, when both variables $i, j$ are scaled by their standard deviations, the indirect effect of a structural uncertainty shock $\varepsilon_j$ on variable $i$, $i \neq j$, is lower than its direct effect on variable $j$. To some extent, these restrictions imply that any of our two uncertainty measures is better than the other one in capturing a structural shock that stems from its own domain – a plausible assumption given the obvious methodological differences between the two indicators. Indeed, despite the inherent statistical overlaps, CISS is a composite indicator designed, and empirically tested (see Hollo et al., 2012), to quantify financial market stress rather than Knightian uncertainty; similarly, EPU is designed to quantify policy uncertainty as reflected in the media and related to government's initiatives, public proposals, or changes in rhetoric and opinions rather than financial stress.

---
[24] To better illustrate this point, not including the country-specific foreign vector $Y_{i,t}^*$ would rise all these cross-sectional correlations to within the 0.2 – 0.4 range.

In our framework, however, we are interested in multi-country spill-overs and so we need to extend the identifying approach to a GVAR setting. The basic idea is that a shock, originated in a specific country of the model, cannot have on-impact effects (and spill-overs) larger than the contemporaneous effect on the domestic variable itself. In other words, the magnitude restrictions approach, for each of the two uncertainty shocks of interest, must extend to the spill-overs as well, rather than being concentrated on the responses within a country. The object of the magnitude restriction, thus, will be the entire column of $S$, for each shock and for each country. Although the extension is trivial from a theoretical point of view, it poses important challenges for its implementation. In this paper, thus, we extend the De Santis and Zimic's algorithm, proposed for small systems (see the Appendix from De Santis and Zimic, 2018), to a GVAR setting. Appendix 3 details the main steps of the algorithm as implemented in our GVAR specification. The challenging point is that the whole set of magnitude restrictions imposed in the GVAR enormously reduces the number of admissible $Q$ matrices among those randomly generated in standard algorithms used in the literature implementing inequality restrictions.[26] As can be seen in Appendix 3, we propose a solution based on block-diagonal $Q$ matrices with perturbations. We solve the identification issue within each country by generating $N$ admissible orthogonal matrixes $Q_1, \dots, Q_N$, one for each of the $N$ countries, and form the admissible block-diagonal orthogonal matrix $Q = diag(Q_1, \dots, Q_N)$. We rotate $Q$ by a *small rotation* matrix $(I - H)(I + H)^{-1}$, with $H$ hemisymmetric, i.e. $\tilde{Q} = Q(I - H)(I + H)^{-1}$, and then check for the magnitude restrictions on the columns of interest of the obtained matrix $S = \Omega_{tr}\tilde{Q}$. This strategy, allows to enormously increase the success rate.

---

[25] This means that the two uncertainty variables are allowed to move contemporaneously in any direction in response to a structural shock, as along as the relative (measured in terms of standard deviations) impact fulfils the respective inequality.

[26] The success rate is in the order of 1 over 10,000, i.e. 0.01%.

We thus contribute to the existing literature by adopting magnitude restrictions in order to identify various types of uncertainty shocks stemming from overlapping sources. In this context, it is important to discuss the advantages of our approach in relation to other structural identification methods available in the broader VAR literature. Firstly, our identification through magnitude restrictions does not impose any time precedence on the two uncertainty variables, like would be the case when applying a standard Cholesky identification (which is just a special case of the identification based on magnitude restrictions as it imposes a zero contemporaneous response of some variables to some shocks).[27] In our case, imposing a time precedence between two uncertainty proxies would be too strong of an assumption, given the complex, dynamic, double causality influences between policy and financial uncertainty. For example, in the particular cases of Greece and Ireland that we discussed in the introduction, the precedence of the shocks is reversed (see Farhi and Tirole, 2017); however, for most other situations that are relevant for empirical analyses, it is not that clear which of the two uncertainty shocks would come first.

Secondly, an alternative identification method based on sign restrictions would require strong theoretical predictions about the transmission mechanisms underlying the two types of uncertainty shocks. This might be hard to achieve when conceptual overlaps are present, particularly in the case of uncertainty where a perfect match between the theoretical notion and its empirical counterpart remains challenging (see discussion in Jurado et al., 2015). Moreover, as noted in Caldara et al., (2016), different uncertainty shocks, despite differences in measurement, can have similar effects on other macroeconomic variables, complicating identification.[28] Thirdly, Bacchiocchi (2017) and Angelini et al., (2019) build on the original "identification through heteroskedasticity" idea proposed in Rigobon (2003) in order to identify uncertainty shocks in a VAR model. While their method is successful in dealing with endogeneity challenges that arise between uncertainty and real or financial variables, it requires that (at least some) structural parameters remain constant over time and across volatility regimes. Fourthly, Caldara et al., (2016) identify the effects of economic uncertainty and financial shocks by employing a penalty function approach, which shares some similarities with our identification approach. In their case, the structural shock should maximize the impulse response of its respective target variable over a pre-defined period. However, although they are able to identify the two structural shocks, they still use a sequential identification due to reverse causality fears. Fifthly, Piffer and Podstawski (2018) use external instruments (e.g. the price of gold) to identify uncertainty shocks. While effective in other applications, applying their approach to our setting would require finding not one, but two distinct instruments (i.e. one for each uncertainty proxy), which is an even greater challenge.

---

[27] Bekaert et al. (2013) estimate a VAR specified in business cycle, monetary policy, risk aversion and expected market volatility, using a Cholesky decomposition (with variables ordered as listed), and a combination of contemporaneous with long-run restrictions. They find that risk aversion decreases more strongly than volatility to a lax monetary shock, with both expected volatility and risk aversion extracted from VIX. Others, like Baker et al. (2016) and Jurado et al. (2015), also employ Choleski decompositions, but use a single uncertainty proxy, not two different ones.
[28] As the required inequality restrictions must be fulfilled only in absolute terms in our case, EPU and CISS are free to either co-move or move in opposite directions, and they might have similar effects on bond spreads.

## 4. RESULTS

### 4.1. Preliminary data analysis

As a preliminary data analysis, we provide arguments for the empirical overlaps existing between policy and financial uncertainty. Tables 2 and 3 below display the correlations between country-specific EPU and CISS indexes, in log terms, computed over the entire sample (for countries where EPU is available), at monthly frequencies. The main challenge to our identification of country-specific shocks rests specifically on these substantial correlations, measured both across as well as within countries.

With the noticeable exception of U.K. and Sweden, almost all correlations between EPU and CISS in Table 2 are positive and statistically significant. For France and Ireland, correlations are slightly weaker when CISS lags EPU. The magnitude of the correlations is higher when EPU lags CISS in case of France, Germany, Netherlands, Spain and Ireland, but lower in case of Italy and Greece. In Table 3, we see that the cross-border contemporaneous correlations among similar types are even higher than the pair-wise correlations displayed in Table 2. This observation highlights the bigger challenge we face in identifying *country-specific* uncertainty shocks within the EU (or the EA), where common components might contribute more in driving uncertainty dynamics than country-specific components.

We caution the readers not to make any causality inference from these correlations, which lack sufficient robustness and sometimes change with the sample size and period. This lack of robustness, instead, should be interpreted as an illustration of the dynamic nature of the spill-overs and interactions between policy (EPU) and financial (CISS) uncertainties, which might amplify or cancel each other, depending on the period, or the nature of the triggering event (source) in a particular country. Once we identify the structural shocks from the reduced-form residuals, we can investigate the overlapping of the structural shocks' time-series with some well-known episodes that marked the recent history of some of the countries under consideration.

**Table 2:** Within-country pair-wise correlations between EPU and CISS indexes

| Country | EPU(t) x CISS(t-2) | EPU(t) x CISS(t-1) | EPU(t) x CISS(t) | EPU(t-1) x CISS(t) | EPU(t-2) x CISS(t) |
|---|---|---|---|---|---|
| **FR** | 0.104 | 0.1485** | 0.194*** | 0.189*** | 0.147** |
|  | (0.15) | (0.043) | (0.007) | (0.009) | (0.045) |
| **DE** | 0.1208 | 0.179** | 0.2478*** | 0.218*** | 0.1637** |
|  | (0.102) | (0.014) | (0.000) | (0.003) | (0.026) |
| **EL** | 0.3345*** | 0.3177*** | 0.3108*** | 0.3078*** | 0.2835*** |
|  | (0.000) | (0.000) | (0.000) | (0.000) | (0.000) |
| **IT** | 0.4638*** | 0.4279*** | 0.4577*** | 0.4009*** | 0.3622*** |
|  | (0.000) | (0.000) | (0.000) | (0.000) | (0.000) |
| **IE** | 0.125* | 0.1118 | 0.1307* | 0.1551** | 0.1457** |
|  | (0.091) | (0.129) | (0.075) | (0.035) | (0.048) |
| **NL** | 0.2355*** | 0.2392*** | 0.287*** | 0.2601*** | 0.2637*** |
|  | (0.001) | (0.001) | (0.000) | (0.000) | (0.000) |
| **ES** | 0.2685*** | 0.3177*** | 0.4008*** | 0.3735*** | 0.328*** |
|  | (0.000) | (0.000) | (0.000) | (0.000) | (0.000) |
| **SE** | -0.2456*** | -0.2019*** | -0.1518** | -0.1526** | -0.1546** |
|  | (0.000) | (0.005) | (0.038) | (0.037) | (0.035) |
| **UK** | 0.0309 | 0.0314 | 0.011 | -0.0319 | -0.0586 |
|  | (0.677) | (0.670) | (0.881) | (0.666) | (0.428) |

Note: The effective sample is: 2003:M01 – 2018:M06. The first rows display the lag/lead structure of the two time-series for which we compute the correlations, with t-1, t-2 and t+1, t+2 denoting 1 and 2 period lags, and leads respectively. Both EPU and CISS time series are in log terms. The p-values are provided in parentheses. The *, ** and *** denote statistical significance at 10%, 5% and 1% respectively.

**Table 3:** Cross-country contemporaneous correlations for EPU and CISS indexes

| Country | FR | DE | EL | IT | IE | NL | ES | SE | PT |
|---|---|---|---|---|---|---|---|---|---|
| | | | | **CISS(t) x CISS(t)** | | | | | |
| **DE** | 0.735*** | | | | | | | | |
| **EL** | 0.486*** | 0.416*** | | | | | | | |
| **IT** | 0.665*** | 0.507*** | 0.774*** | | | | | | |
| **IE** | 0.715*** | 0.557*** | 0.420*** | 0.597*** | | | | | |
| **NL** | 0.723*** | 0.733*** | 0.401*** | 0.487*** | 0.741*** | | | | |
| **ES** | 0.766*** | 0.577*** | 0.699*** | 0.818*** | 0.703*** | 0.604*** | | | |
| **SE** | 0.627*** | 0.684*** | 0.255*** | 0.408*** | 0.694*** | 0.634*** | 0.494*** | | |
| **PT** | 0.679*** | 0.405*** | 0.708*** | 0.854*** | 0.625*** | 0.467*** | 0.857*** | 0.351*** | |
| **UK** | 0.561*** | 0.578*** | 0.390*** | 0.520*** | 0.710*** | 0.578*** | 0.541*** | 0.702*** | 0.469*** |

| Country | FR | DE | EL | IT | IE | NL | ES | SE |
|---|---|---|---|---|---|---|---|---|
| | | | | EPU(t) x EPU(t) | | | | |
| DE | 0.776*** | | | | | | | |
| EL | 0.665*** | 0.617*** | | | | | | |
| IT | 0.538*** | 0.479*** | 0.453*** | | | | | |
| IE | 0.570*** | 0.522*** | 0.390*** | 0.395*** | | | | |
| NL | 0.280*** | 0.290*** | 0.201*** | 0.504*** | 0.168** | | | |
| ES | 0.636*** | 0.680*** | 0.640*** | 0.563*** | 0.472*** | 0.328*** | | |
| SE | 0.681*** | 0.671*** | 0.559*** | 0.448*** | 0.460*** | 0.277*** | 0.588*** | |
| UK | 0.821*** | 0.758*** | 0.655*** | 0.432*** | 0.651*** | 0.197*** | 0.645*** | 0.690*** |

Note: The effective sample is: 2003:M01 – 2018:M06. Both EPU and CISS time series are in log terms. The *, ** and *** denote statistical significance at 10%, 5% and 1% respectively.

## 4.2. Estimation of the GVAR model and its main results

With all variables expressed in log terms (except for spreads), we estimate the model directly in levels, allowing an easy interpretation of its impulse responses, which provide us with the main insights. Sims et al., (1990) recommend against differencing even in the presence of unit roots, arguing that the goal of the analysis should be to determine the interactions between variables. They show that the VAR specified in levels delivers consistent estimates, even in the presence of stochastic trends and cointegration. Elliot (1998) further shows theoretically that imposing cointegration for near unit root variables can lead to large distortions. We do not estimate cointegrating relations, nor include time trends and error correction terms, also because our short sample and small set of variables would preclude a robust identification of these long-term relationships.[29]

Our sample includes more than 15 years of monthly observations. The main trade-off we are facing in the estimation is between model parsimony and its statistical properties (e.g. stability, residual tests). Kapetanios et al. (2007) notice that the quality of a VAR approximation to the true model depends on both the number of variables and the lag order; as the GVAR includes more variables than a normal VAR (i.e. both domestic and foreign variables in each country-specific model), small lag orders are regularly employed. We set a small lag order for domestic variables, i.e. $pi = 2$ for all countries, because it helps eliminate most residual autocorrelation (or serial dependence) and preserves a parsimonious specification (i.e. setting a higher lag order would further reduce autocorrelation). Regarding the maximum lag allowed for foreign variables, $qi$, a smaller lag order than $pi$ is to be preferred because markets in general react rapidly to foreign influences (e.g. media news, financial volatility boosts). Unfortunately, setting $qi = 0$ for all countries does not guarantee model stability in all its different specifications (including robustness checks). Instead, allowing for $qi = 1$ in very few cases, particularly for DE and UK (and some of their close neighbours, like AT, BE, IE, DK), appears the easiest fix to this stability problem and, in addition, maintains model parsimony. In the majority of cases, the inclusion of country-specific foreign variables is supported by F-tests, where the null restricts the $C_{i,j}$ coefficients in equation (1) to zero, in each country $i$.[30] As for the ECB model, we chose a parsimonious specification similar to the overall GVAR, setting the lag orders as $px = 2$ and $qx = 1$.

---

[29] Both theory and empirical studies provide evidence that European sovereign spreads are cointegrated with fundamentals (e.g. fiscal proxies, economic and financial proxies), which are omitted from our estimated GVAR (see also De Santis, 2020). Expanding this GVAR specification could be an interesting avenue for future research.

Figure 1 depicts the estimated residual autocorrelation and the eigenvalues of the estimated GVAR. All residual autocorrelations lie within, or very close to, the confidence bands (set at $\pm 2$ standard deviations); given the large number of variables and the inherent over-parameterization of the GVAR, a small number of larger autocorrelations are to be expected. Despite some inherent persistency, all model eigenvalues are strictly less than one.

**Figure 1**: GVAR specification checks

Panel A: Residual autocorrelation   Panel B: Eigenvalues of GVAR

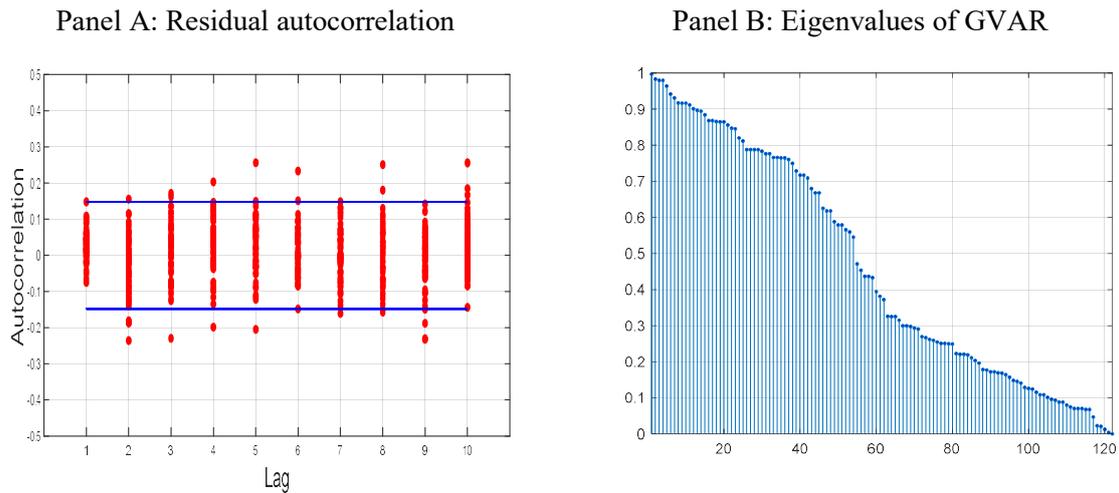

Note: Panel A plots the residual autocorrelations for all GVAR variables and all country-specific models, as a function of the serial lag. Panel B plots the GVAR eigenvalues.

To derive our insights, we rely on the models' impulse response functions (IRFs) to the two uncertainty shocks identified through magnitude restrictions. To gauge statistical significance of the IRFs, we use bootstrapped 68% confidence intervals[31] based on 500 replications, allowing for a maximum of 100 draws of the orthogonal matrix $\tilde{Q}$ at each replication; the success rate is around 25%. Table 4 below conveniently summarises the main results to a series of uncertainty shocks originating in IT, ES, EL, IE, and FR – a relevant group that allows us to draw interesting insights. The first 4 countries in this list were seen (at some particular moments) as the most vulnerable EA members, while France makes for an interesting case as it had to weather a series of recent uncertainty shocks, mostly originating in the policy realm. More detailed plots for the 5 countries are given in Appendix 4; the complete set of IRF results can be obtained from the authors upon request.

---

[30] Results of the F-tests are available upon request from the authors.
[31] Burriel and Galesi (2018) and Anaya et al. (2017) also use 68% confidence intervals in GVAR specifications, which are known to deliver wider confidence bands due to over-parameterization.

Table 4: Summary findings based on IRFs to identified uncertainty shocks

| Shock origin | Observed responses | CISS responses to a CISS shock | EPU responses to a CISS shock | CISS responses to a EPU shock | EPU responses to a EPU shock |
|---|---|---|---|---|---|
| **Italy** | domestic | Significant up to 9 months | Insignificant | Insignificant | Significant up to 6 months |
| | cross-border | Significant up to 9-18 months (depending on the country) | Significant up to 3-18 months (depending on the country) | Insignificant, except for PT | Significant up to 3-12 months (depending on the country) |
| **Spain** | domestic | Significant up to 9 months | Insignificant | Insignificant | Significant up to 9 months |
| | cross-border | Significant up to 9-18 months (depending on the country) | Weakly significant in few countries | Insignificant, except for PT | Significant up to 6-12 months (depending on the country) |
| **Greece** | domestic | Significant, up to 12 months | Significant up to 9 months | Insignificant | Significant up to 9 months |
| | cross-border | Significant up to 6 months in most countries, except DE, NL | Insignificant | Significant up to 6-9 months, only in EA countries | Significant up to 6-9 months (depending on the country) |
| **Ireland** | domestic | Significant up to 12 months | Significant up to 9 months | Insignificant | Insignificant beyond impact period |
| | cross-border | Significant up to 9-18 months (depending on the country) | Significant up to 6-18 months, except for FR | Insignificant | Insignificant beyond the impact period and only in IT, ES, UK |
| **France** | domestic | Significant up to 9 months | Insignificant | Insignificant | Significant up to 9 months |
| | cross-border | Significant up to 9-18 months (depending on the country) | Significant up to 3-24 months in few countries like IT, NL, IE, SE | Weakly significant in few countries like DE, IT, PT, EL, PL | Significant up to 6-9 months (depending on the country) |

Note: The table displays a summary description of the IRFs to the two identified uncertainty shocks. Statistical significance is assessed based on the bootstrapped 68% confidence intervals derived from 500 replications, with a maximum of 100 maximum draws of the orthogonal matrix $\tilde{Q}$ at each bootstrap replication.

Three main findings emerge from Table 4. Firstly, there are substantial and persistent cross-border effects generated by domestic EPU and CISS shocks. In some cases, these effects are even more persistent than the domestic responses, for which the initial magnitude has been restricted (by the identification strategy) to be the highest. For example, Italian EPU shocks generate more persistent responses in French and British policy circles rather than in Italy; instead, Italian CISS shocks generate more persistent responses in the Baltics group. Likewise, Spanish and Irish CISS shocks generate more persistent responses in some Central European (e.g. DE, HU) and Nordic (e.g. SE, BA) countries, pointing at underlying vulnerabilities or high exposure towards the origin countries (mostly in the case of Ireland).

Secondly, we see important overlaps between policy and financial realms as revealed by the substantial but asymmetric interactions (i.e. cross-influences) we observe between the two uncertainty proxies. We find cases where domestic EPU responds to domestic CISS shocks in Greece and Ireland, but no case where CISS responds to EPU shocks within the same country. Except for Greece, most foreign EPUs respond significantly to domestic CISS shocks as well. On the contrary, foreign CISS responses to domestic EPU shocks, even when statistically significant, are only weak and slow. In other words, it is more likely that policy uncertainty responds to shocks in financial uncertainty, both within as well as across countries, and less likely that financial uncertainty reacts to policy uncertainty shocks. This asymmetry in responses is remarkable given the symmetric treatment of the two uncertainty proxies in the specification of EU countries' VARX models. The result, however, is in line with much of the existing evidence on the importance of financial frictions for real (macro)economic dynamics and therefore for policy stability (Allen et al., 2011).

A third finding, not summarised in Table 4 to save space, but easily revealed by the figures displayed in Appendix 4 (panels C), refers to sovereign spreads' reactions to uncertainty shocks. According to these IRFs, reactions to CISS shocks illustrate a rapid overshooting followed by a slow undershooting of the initial spreads' level within a two-year period; spreads' reactions to EPU shocks instead are less likely to be statistically significant, but still positive, and returning slowly to the initial level. In other words, higher financial (but not policy) uncertainty is likely to lead to higher volatility in sovereign spreads over the medium term.

The above IRFs provide a summary of the total estimated effects of the shocks, but do not really reveal the importance of cross-border linkages. To this end, we proceed by decomposing the IRFs into: (i) cross-border spill-overs and (ii) direct effects (see also Burriel and Galesi, 2018). The direct effects are obtained from a restricted GVAR specification where we shut down the cross-country linkages that arise through the presence of foreign variables, i.e. we set the $C_{i,j}$ coefficients in equation (1) to zero in the restricted GVAR specification; the direct effects are therefore an estimation of the impact that arises due to the interaction of the model's endogenous variables with the common variables.[32] Necessarily, the spill-overs will be computed as a difference between the restricted and unrestricted GVAR responses.

In Appendix 4, in each figure's panel D, we present such a decomposition for the median peak IRFs, computed over the first 6 months (when peak responses are the most likely). What can be easily revealed in these plots is the sizable contribution of spill-overs in determining the foreign uncertainty responses to domestic uncertainty shocks, but of a different type; instead, spill-overs' contribution is relatively minor in the case of foreign responses to domestic uncertainty shocks of the same type. In other words, once an uncertainty shock crosses the national boundaries, its true nature and origin (i.e. either from policy or financial realms) can become blurred – creating thus a cross-border uncertainty-amplifying mechanism. As an example, French EPU shocks generate peak responses in German and Portuguese CISS that are determined to a larger extent by spill-overs rather than by direct effects; this is not the case for foreign CISS responses to a French CISS shock or for foreign EPU responses to a French EPU shock (for this example see panel D in Figure A4.5, Appendix 4). To make our point clearer, notice that it is precisely because of spill-overs' sizable contribution that Italian CISS responses are higher than Italian EPU responses to the same French EPU shock – pointing thus to a possible confusion abroad regarding the true nature of a French shock. Clearly, the size of such spill-overs comes from the interaction of all model endogenous variables, including sovereign spreads, whose role in facilitating the transmission of uncertainty shocks reflects international portfolios' rebalancing dynamics.

Such an amplification mechanism for uncertainty shocks also appears to be novel to the existing literature, mostly because models with a strong international dimension and featuring overlapping and/or multiple uncertainty sources are rather rare. Some ingredients of this mechanism can be searched for in theoretical models featuring cross-border information asymmetries we discussed in section 2. In King and Wadhwani (1990) or Kodres and Pritsker (2002) uninformed investors cannot correctly differentiate local idiosyncratic shocks from common shocks that lead to portfolio rebalancing. While a new theory can better discipline our reasoning, a similar argument can be used here if, for example, one simply assumes that policy uncertainty generates idiosyncratic shocks, and financial uncertainty generates common shocks (this would be in line with evidence of deep financial integration, but incomplete EU institutional integration). Our results though show there is a richer dynamics in the data.

### 4.3. ECB policy reactions to uncertainty shocks

One of the main questions we raised in the introduction relates to the ECB role in filling the leadership gap within the EU institutional and governance structure. From an empirical perspective, as long as the

---

[32] We note that the direct effects, in the large majority of cases, have the most important contribution to the median IRFs, highlighting thus, among other things, the key role of VIX as a proxy for global uncertainty or risk.

ECB policy proxies are included in the EA country models, their relevance can be easily validated. Table 5 below presents, by country $i$ and by equation (or variable), the F-statistics of the null hypothesis that in equation (4) we have $D_{i,j} = 0$, jointly for all $j = [0, qi]$ in country $i$. Results are rather mixed, illustrating that ECB had varying degrees of influence (from a statistical perspective) on EA countries; for Italy and Greece the impact of ECB proxies falls mostly on policy uncertainty, while for France, Portugal, and Central Europe it falls on financial uncertainty; sovereign spreads are instead affected by ECB in the cases of Austria, the Netherlands, Greece, Spain, Portugal and Slovenia.

**Table 5:** F-tests of the joint null that ECB policies had no influence on a EA country/equation

| Country: | Equation: Spread | Equation: CISS | Equation: EPU | crit. (5%) |
|---|---|---|---|---|
| Austria | 3.6655* | 0.8493 | | 2.1532 |
| Belgium | 1.0004 | 0.9094 | | 2.1532 |
| Finland | 0.5768 | 0.5293 | | 2.6565 |
| France | 1.3005 | 4.1920* | 0.6772 | 2.6571 |
| Germany | | 0.7373 | 1.5160 | 2.1532 |
| Italy | 1.3990 | 0.8465 | 4.8750* | 2.6571 |
| Netherlands | 3.5612* | 2.3054 | 0.7947 | 2.6571 |
| Spain | 2.7287* | 2.5343 | 0.0899 | 2.6571 |
| Greece | 0.2254* | 1.4442 | 4.5651* | 2.6571 |
| Ireland | 0.3269 | 1.6760 | 0.5311 | 2.1539 |
| Portugal | 3.4224* | 6.7346* | | 2.6565 |
| Luxemburg | 1.2558 | 0.4757 | | 2.1532 |
| Slovakia | 1.4932 | 6.0780* | | 2.6565 |
| Slovenia | 9.6018* | 3.5474* | | 2.6565 |
| Baltics | 1.0484 | 0.1417 | | 2.6565 |

Note: The table displays the F-statistics computed by restricting the coefficients of the three ECB proxy variables to be exactly zero (see equation 4) in the country/equation indicated on the first column/row of the table. The degrees of freedom associated with the F-statistics vary, depending on the number of domestic variables and the lag structure of foreign variables in each VARX model. The * denotes cases where the F-statistics is higher than the critical value, and therefore the null can be rejected at a 5% significance level.

An analysis of the IRFs associated with ECB policy responses to uncertainty shocks provides a complementary perspective to our previous insights; see Appendix 5 for more details. Concentrating on the same subset of 5 countries as in the previous section, we find that ECB would adjust its $CMP$ in line with expectations, i.e. ECB MRO drops in response to (at least one) uncertainty shocks from either IT, ES, EL, IE or FR. With the exception of Greece, ECB always reduces its MRO more abruptly and swiftly (i.e. within 2 months) in response to CISS shocks, than in response to EPU shocks. ECB is also quick in increasing liquidity in response to CISS shocks from all vulnerable countries, again except

Greece. Meanwhile, its liquidity responses to EPU shocks are more sluggish, and in the case of Spain even statistically insignificant. Regarding its *UMP* reactions, the IRFs show that the ECB balance sheet responds statistically significant only to EPU shocks, but not to CISS shocks; in the case of Spanish shocks, however, we find no statistically significant reactions to any of the two shocks. In case of Italian CISS shocks, we also see a weak and short-lived reaction in ECB *UMP* proxy.

These results require a more detailed discussion. For the ECB, setting up and enabling its *UMP* toolbox has required more effort and political compromise[33] than deploying its *CMP* or regular liquidity facilities – most of which are rather standard tools, scheduled in advance and/or automatically triggered by specific market events. In line with much of the prior literature, our proxy for *UMP* is the annual change in ECB balance sheet, reflecting the size of the additional financing that gets channelled through the banking system. Central banks' balance sheets have attracted much research attention, with Gambacorta et al., (2014) being the first to popularise the use of central bank's balance sheet as a proxy for *UMP*. Starting with Fratzscher et al. (2016), there is also a growing list of studies focusing on ECB's unconventional monetary policy effects; see among many others Moder (2017), Burriel and Galesi (2018), Boeckx et al., (2017), Kucharcukova et al. (2016), Koijen et al., (2018).[34]

Obviously, policy innovation with respect to its *UMP* toolkit has required significant leadership from ECB, who even had to face legal challenges in defending its approach.[35] The fact that we find statistically significant *UMP* responses only in reaction to EPU shocks is therefore remarkable. Accordingly, our new results in this section suggest a more pro-active ECB stance towards policy uncertainty shocks, and a more accommodative (i.e. passive) stance towards financial uncertainty shocks where less policy innovation was required. De Grauwe and Ji (2013) also advocate for a more active ECB role in counteracting self-fulfilling crises driven by investors' fears, not fundamentals, claiming that EA fragility (as perceived by investors) stems from the lack of a "lender of last resort" for both banks and sovereigns. Moreover, De Grauwe and Ji (2013) underline the role of information frictions in an EA characterised by advanced financial and economic integration, but not enough political (including fiscal policy) integration. In a similar vein, and drawing on the theoretical literature discussed in section 2, we can posit that ECB has tried to dampen policy uncertainty shocks in order to prevent a financial market "segmentation equilibrium" in which information frictions would harden along the existing national border lines (see Freixas and Holthausen 2004).

---

[33] ECB unconventional monetary policy tools include various non-standard liquidity provision (LTROs) and several asset purchase programs. For example, ECB conducted two Covered Bond Purchase Programs (CBPPs) between June 2009 and October 2012; between 2010 and 2012, ECB bought government bonds on the secondary market through its Securities Markets Program (SMP). After November 2014, ECB conducted an asset-backed securities purchase program (ABSPP) and a third CBPP.

[34] As noted in Boeckx et al., (2017), identification of unconventional policy shocks can be challenging if it does not sufficiently distinguish between policy- and demand-driven shocks. Given ECB strategy of fixed-rate tenders with full allotment, isolating the exogenous shifts in ECB's balance sheet becomes key for proper identification. As long as some relevant transmission mechanisms and indicators (e.g. real business cycle indicators) are missing from our model, we do not attempt to identify conventional, nor unconventional monetary policy shocks.

[35] An example is the recent decision of the German Constitutional Court on ECB's Public Sector Purchase Programme. See: https://www.ecb.europa.eu/press/pr/date/2020/html/ecb.pr200505~00a09107a9.en.html.

### 4.4. Back-testing

De Santis and Zimic (2018) admit that their magnitude restrictions are inspired by event study techniques, which require a good understanding of the historical patterns present in the time-series being modelled. As already mentioned, magnitude restrictions can provide a mathematical approach to identification in a VAR including some highly correlated time-series. While this approach guarantees that, on a particular time moment, the identified shock has the largest magnitude among all the other shocks, there is no guarantee that the shock has any real, meaningful interpretation. This section addresses this issue and provides evidence on the suitability of using magnitude restrictions in our case.

We draw on multiple media sources to categorise a set of unique country-specific events that stand as outliers in the time-series of the identified uncertainty shocks (Appendix 6 provides an overview on the complete distribution of these shocks, along with their prevalence in different countries). Figure 2 below illustrates some major events that shaped the recent history of some European countries. To facilitate interpretation, for each event we plot the two uncertainty shocks (in the leftmost panel for each event), as well as a comparison of the time profile for the same-type uncertainty shocks in all remaining countries for which we perform the identification (see the middle two panels for each event).

**Figure 2**: Overlap of identified uncertainty shocks and historical major country-specific events

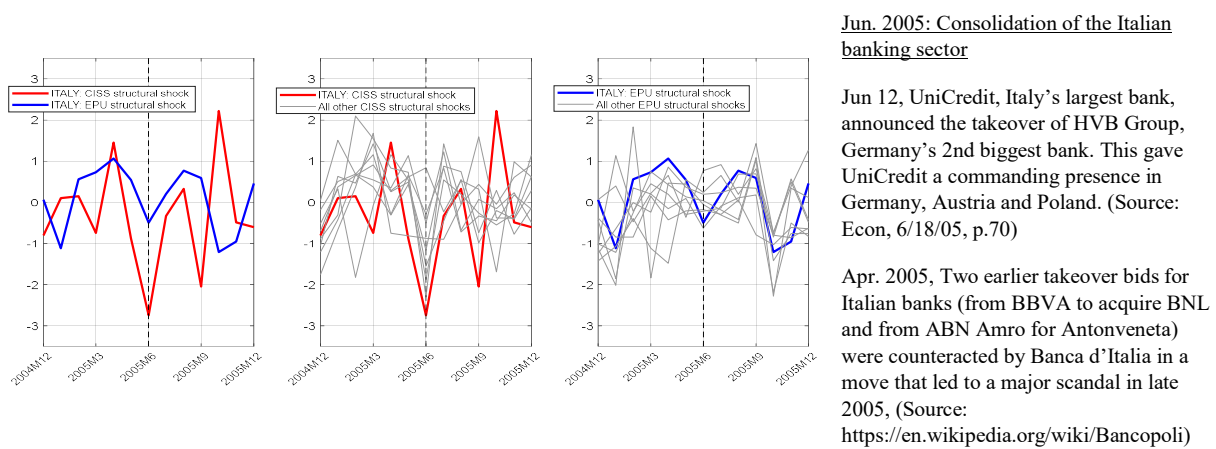

Jun. 2005: Consolidation of the Italian banking sector

Jun 12, UniCredit, Italy's largest bank, announced the takeover of HVB Group, Germany's 2nd biggest bank. This gave UniCredit a commanding presence in Germany, Austria and Poland. (Source: Econ, 6/18/05, p.70)

Apr. 2005, Two earlier takeover bids for Italian banks (from BBVA to acquire BNL and from ABN Amro for Antonveneta) were counteracted by Banca d'Italia in a move that led to a major scandal in late 2005, (Source: https://en.wikipedia.org/wiki/Bancopoli)

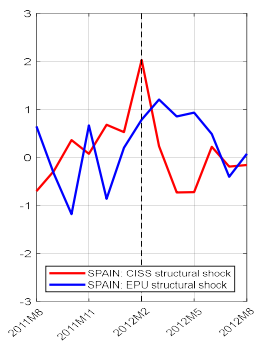 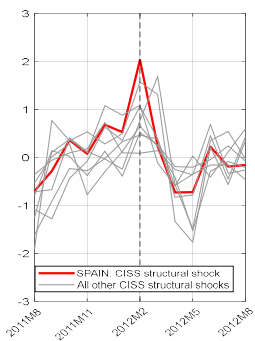 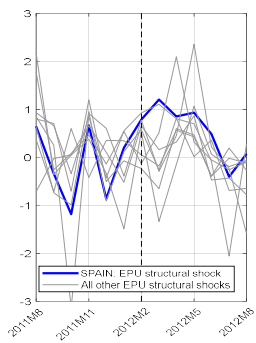

Feb. 2012: Lax attempts to reform bring Spanish banks under pressure

Royal Decree-Law 2/2012 on balance sheet clean-up of the financial sector required provisions to be recorded by 31 December 2012 for foreclosed real-estate development and construction loans and assets relating to credit institutions' business in Spain as at 31 December 2011. (Source: Banco de Espania, 51 Report on Banking Supervision in Spain, 2012)

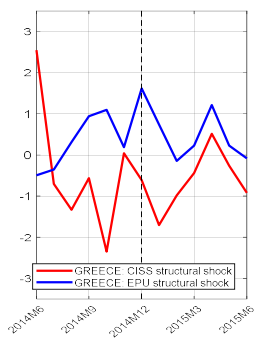 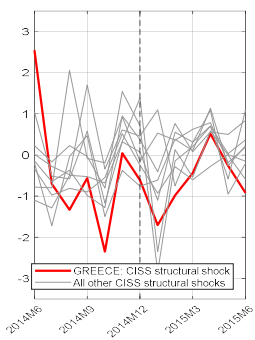 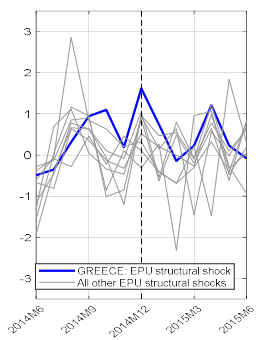

Dec. 2014: Snap elections in Greece

Snap elections are scheduled for January 2015 due to the failure of the Greek parliament to elect a new president in Dec. 2014. (Source: https://en.wikipedia.org/wiki/January_2015_Greek_legislative_election)

Dec., 29, Due to rising political uncertainty IMF suspends all scheduled remaining financial aid to Greece under its bailout programme (Source: Kathimerini, 12/29/14)

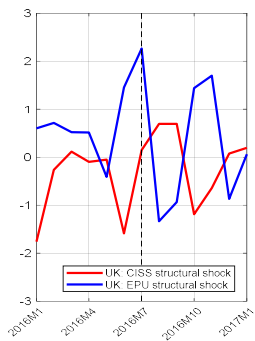 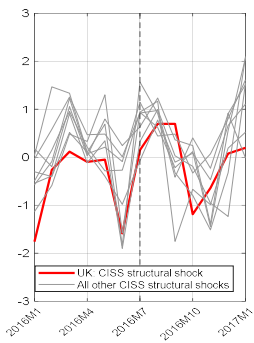 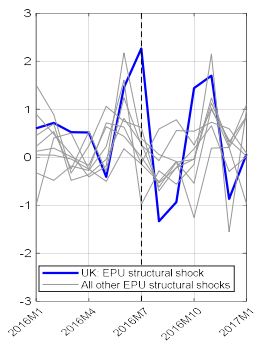

Jul. 2016: Brexit referendum

Jun 23, Britons voted to exit the European Union. Scotland voted decisively to stay in the EU by 62 to 38 percent in the referendum, putting it at odds with the United Kingdom as a whole, which voted 52-48 percent in favor of an exit from the EU, or Brexit. 17.4 million people voted to leave. (Source: Reuters, 6/23/16; Reuters, 6/24/16; Econ, 7/2/16, p.14)

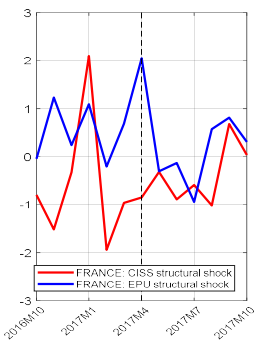 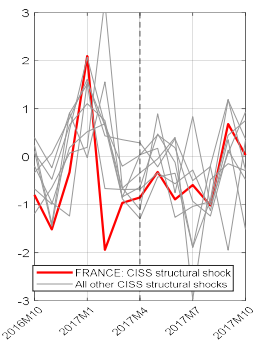 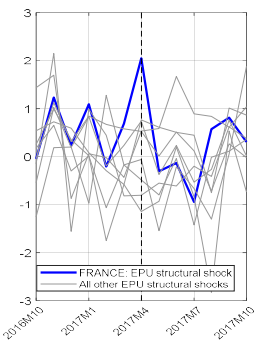

Apr. 2017: Presidential elections in France

No candidate wins a majority in the first round of elections held on 23 April. A run-off was held between the top two candidates, Emmanuel Macron of En Marche! and Marine Le Pen of the National Front (FN). (Source: https://en.wikipedia.org/wiki/2017_French_presidential_election)

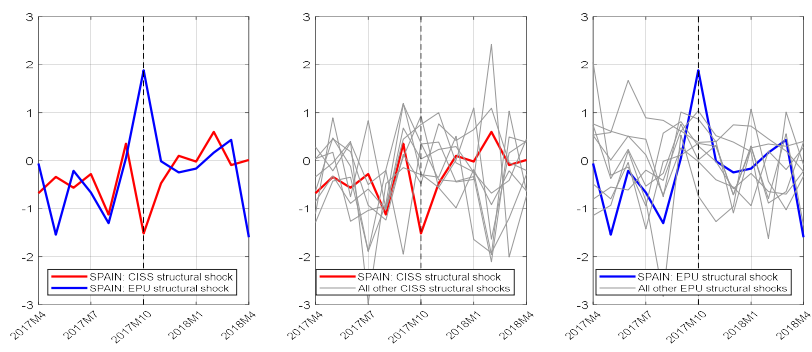

| | Oct. 2017: Catalonia's independence referendum |
|---|---|
| | Oct., 11, Spanish PM Mariano Rajoy threatened to impose direct rule on Catalonia following its disputed independence referendum. (Source: AFP, 10/11/17; AP, 10/11/17) |
| | Oct., 17, Spain's top court officially ruled that Catalonia's disputed independence referendum was illegal. (Source: AP, 10/17/17; Reuters, 10/17/17) |

Note: The date of the main historical event indicated in text on the right side is also highlighted with a vertical dash line in the associated graphs. Uncertainty shocks are identified in the following countries: DE, IT, ES, FR, EL, SE, IE, NL and UK. News and headlines are borrowed from the following online repositories: www.timelines.ws, https://en.wikipedia.org and https://en.wikipedia.org/wiki/Portal:Current_events, although the original source might be different, as indicated in the text.

### 4.5. Robustness checks

Three different robustness checks are performed in order to validate our main findings. We next provide the main technical details behind their implementation and describe the results.

As a first robustness check, we add a measure of global liquidity risk, i.e. the TED spread, which is the spread between the 3-Month LIBOR based on US dollars and the 3-Month Treasury Bills (see Brunnermeier, 2009). Although our GVAR already includes a liquidity proxy relevant for EA markets (i.e. the spread between EONIA and the main ECB policy rate), the US dollar-denominated funding costs of European banks play a key role within the literature on global financial cycles[36] (see Rey, 2015; Bruno and Shin, 2014). When uncertainty raises, banks charge themselves higher interest rates for uncollateralised loans (i.e. LIBOR rate is the reference rate for interbank lending) compared to the yield of risk-free US Treasuries, and therefore the TED spread is actually a global liquidity proxy. The cost of US dollar funding has been a central element of the policy reactions during the peaks of the financial crisis of 2007/2008. All major central banks, including ECB, set up direct currency swap lines with the US Federal Reserve System, precisely to alleviate pressures from the US dollar funding.[37] By adding the TED spread as an endogenous variable to the US model, we account for changes in global liquidity and US dollar funding, providing a consistency check to our main findings from the previous sections. We find that the main results from sections 4.2 and 4.3 remain qualitatively unchanged.

---

[36] This is because the US dollar is the world's main reserve currency. Similar to the global financial cycle literature, the international bank lending channel, exposed in Schmidt, Caccavaio, Carpinelli and Marinelli (2018), highlights the importance of US dollar funding costs on lending in Europe, particularly in France and Italy.
[37] See https://www.federalreserve.gov/monetarypolicy/bst_liquidityswaps.htm.

As a second robustness check, we compute the sovereign spreads against the 10 year U.S. sovereign yield, which is the global benchmark, rather than against Germany, which is the European benchmark. The only technical change required in the model specification given by equation (2) is that the German VARX now includes the sovereign spreads against US, while the US VARX includes only EPU and VIX as endogenous variables. The main findings are again qualitatively unchanged.

In a third robustness check, we re-estimate the GVAR with a different weighting matrix based on BIS Locational Banking Statistics (LBS) data. Appendix 2 provides more details on the constructions of weights in this case. The GVAR estimated in Eickmeier and Ng (2015) fits the data better when using weights based on BIS LBS banking exposures for financial variables (along with trade weights for their model's real variables). Such weights based on BIS LBS data are also employed in Bicu and Candelon, (2013), Feldkircher and Huber (2016) among others. Yet, capital flows driving bank cross-border exposures are generally more volatile than flows driving portfolio exposures according to balance of payments statistics, and our statistical evidence points to the same result. Despite some important differences in weighting between IMF CPIS data and BIS LBS data[38], estimating the GVAR with weights based on the latter dataset delivers qualitatively similar results.

## 5. CONCLUSIONS

Given the complex intertwining between policy and financial realms in Europe, and in the Euro Area in particular, separating between these two overlapping sources of uncertainty can become empirically challenging. Large strands of the relevant literature concentrate on the sovereign-bank nexus, but in a multi-country setting this approach could be of limited use, mostly due to the incomplete nature of the European integration and institutional governance. Following De Santis and Zimic (2018), we adopt and extend their identification method based on magnitude restrictions in order to, first, separate between these two uncertainty sources and, second, evaluate the spill-over potential of country-specific shocks. After exposing the main advantages of our identification approach, we perform a back-testing exercise in which we find that our uncertainty shocks can match the dates of some remarkable events that marked the recent history of the European project.

In terms of empirical strategy, we employ a GVAR specification that can efficiently summarise the cross-sectional and time-series dynamics of our multi-country dataset in order to investigate country-specific shocks. The empirical findings confirm there are statistically significant and persistent effects from both financial- and policy-driven uncertainty shocks. Besides domestic effects, we find that once an uncertainty shock crosses the national boundaries, its true nature and origin might get blurred, creating therefore substantial spill-overs and revealing an uncertainty-amplification mechanism that is also novel in the literature. All our empirical findings withstand the robustness checks.

---

[38] The IMF CPIS data show that most countries in our sample have over-weighted exposures towards US, UK and LU. Instead, according to BIS LBS data, most countries have over-weighted exposures towards UK and LU.

In terms of policy responses, our results suggest that ECB adopted a pro-active stance towards policy uncertainty shocks – an idea for which the variety and range of ECB unconventional policy measures can be offered as proof. Instead, a more passive or accommodative stance was adopted with respect to financial uncertainty shocks, for which conventional monetary policy and liquidity measures would work best. As the European market seems to be gyrating towards either more integration or more fragmentation after each passing crisis, we posit that ECB has managed to fill a key leadership vacuum when reacting to country-specific policy uncertainty shocks.

**ACKNOWLEDGEMENTS**

We are grateful to Elena Beccalli, Efrem Castelnuovo, Giovanni Caggiano, Giulio Palomba, Giovanni Pellegrino and Eduardo Rossi for their comments and suggestions on earlier versions of this paper. We also thank Alessandro Galesi for advice on how to adapt some of the Matlab codes running under the GVAR package, available from https://sites.google.com/site/gvarmodelling/home. Obviously, all remaining errors are our own.

**Appendix 1: Data description, sources and definitions**

**CISS** – Composite Indicator for Systemic Risk. Frequency: monthly averages. Transformation: natural logarithm. Adjustment: seasonally adjusted using X-12 procedure. Source: ECB warehouse (https://sdw.ecb.europa.eu/browse.do?node=9689686). See Hollo et al. (2012) for the construction methodology. For Baltics, we compute the average of CISS indexes for all three countries.

**EPU** – economic policy uncertainty index, computed by Baker et al. (2016). Transformation: natural logarithm. Adjustment: seasonally adjusted using X-12 procedure. Source: data and methodology available from www.policyuncertainty.com.

**VIX** – the Chicago Board Options Exchange (CBOE) Volatility Index. Frequency: monthly averages. Transformation: natural logarithm. Source: Federal Reserve Bank of St. Louis database.

**Spread** – the difference between 10-year sovereign bond yields for each country and Germany (or US); data on 10-year sovereign bond yields is not available for Turkey, for which we use its 5-year sovereign bond yields. Transformation: before computing the spreads, we apply the following transformation of the yields: $yield^{adjusted} = \frac{1}{12} * \ln(1 + \frac{yield}{100})$ to smooth spikes in the time-series. Source: Eurostat.

**Main Refinancing Operations (MRO) rate** – is the short term interest rate at which ECB provides the bulk of liquidity to the banking system of the Euro Area.[39] Source: ECB warehouse.

**EONIA** – is the Euro Overnight Index average or the Euro Interbank Offered Rate defined as the weighted rate for the overnight maturity, calculated by collecting data on unsecured overnight lending in the EA provided by banks belonging to the EONIA panel.[40] Frequency: monthly averages. Transformation: $yield^{adjusted} = \frac{1}{12} * \ln(1 + \frac{yield}{100})$. Source: ECB warehouse. The liquidity proxy used in the GVAR is the difference between EONIA and the Main Refinancing Operations rate.

**ECB assets** – defined as central bank assets for Euro Area (11-19 Countries). Frequency: monthly, end of month. Transformation: natural logarithm. Source: Federal Reserve Bank of St. Louis database. The UMP proxy used in the GVAR is the annual growth rate of the natural logarithm of ECB assets.

**TED spreads** – defined as the spread between the 3-Month LIBOR based on US dollars and the 3-Month US Treasury Bills. Frequency: monthly averages. Transformation: none. Source: Federal Reserve Bank of St. Louis database.

---

[39] See https://www.ecb.europa.eu/stats/policy_and_exchange_rates/key_ecb_interest_rates/html/index.en.html
[40] See also the conclusions of the public consultation on euro risk-free rates at
https://www.ecb.europa.eu/paym/pdf/cons/euro_risk-free_rates/ecb.consultation_details_201905.en.pdf.

**Appendix 2: GVAR weighting matrixes**

Figure A2.1 below displays the GVAR weighting matrix, $W$, computed based on IMF CPIS data. Weights reflect portfolio allocations from countries mentioned on rows towards countries on mentioned on columns (country labels are according to Table 1 in the main text). The colour of each cell indicates the share of country's portfolio allocation towards other countries, based on the scale displayed on the right of the figure. Each row sums to 1, as countries on the column represent the entire investable universe for the country specified at the start of each row.

**Figure A2.1**:

IMF CPIS weights

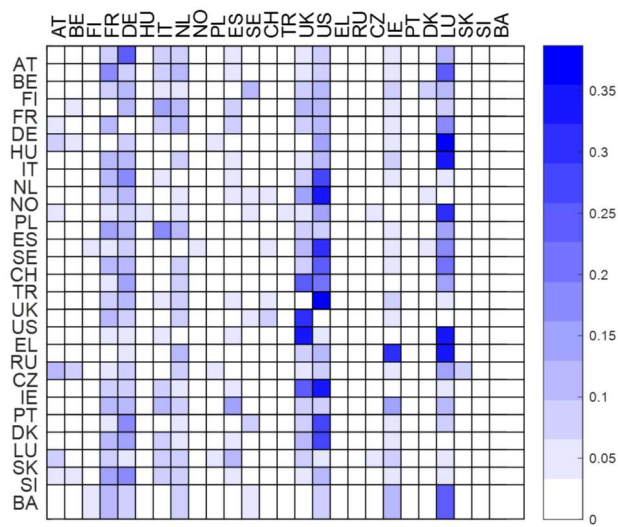

Figure A2.2 below displays the weighting matrix used as a robustness check in section 4.5, based on data from BIS Locational Banking Statistics, tables A6.2.[41] These tables contain data on cross-border positions in mil. USD, by counterparty's country of residency, and by location of the reporting bank. Since not all 28 countries (i.e. 25 individual countries and the 3 Baltics) are reporting to BIS, cross-border positions for banks located in other countries are only indirectly available as the reverse balance sheet positions of banks located in BIS reporting countries; for example, outward claims of banks located in Poland can be inferred as inward liabilities of banks located in BIS reporting countries with Polish resident banks as their counterparties. Moreover, for banks located in BIS reporting countries, there might be some differences between what banks from country X reports as outward claims in country Y, and what banks from country Y reports as inward liabilities from country X. To mitigate the impact of such inconsistencies, we average between (outward) claims and (inward) liabilities for all country pairs, and use bank-to-all sectors rather than just bank-to-bank positions. Further to reduce the impact of time-variation, we average the end of year (4$^{th}$ quarter) exposures over a 7-year period from 2010 to 2016. The colour of each cell indicates the share of country's outward exposures (i.e. claims) towards other countries, based on the scale displayed on the right of the figure. Each row sums to 1, as countries on the column represent the entire investable universe for the country specified at the start of each row.

---

[41] See https://stats.bis.org/statx/toc/LBS.html.

**Figure A2.2**:

BIS LBS weights

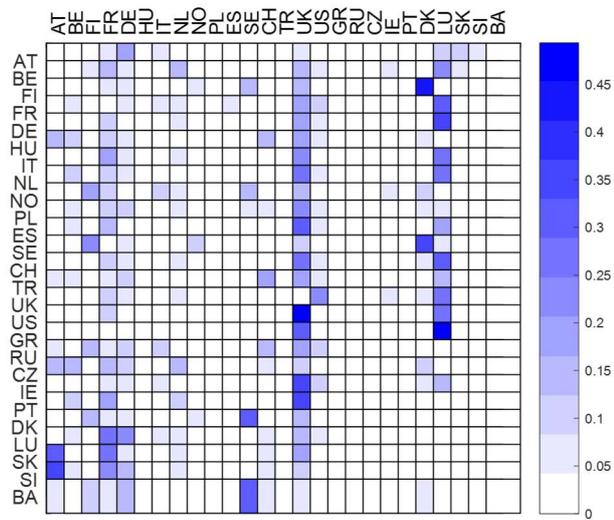

**Appendix 3: The algorithm used for structural identification**

The algorithm is implemented after the estimation of the reduced-form model given in equation (3). To facilitate notation, each country for which we perform the identification is ordered first (i.e. model solution matrixes are reshuffled accordingly). The algorithm is implemented in Matlab, within the estimation codes of the GVAR package, and consists in the following steps:

1. Bootstrap the reduced-form GVAR model given in equation (3) to obtain the variance-covariance matrix of reduced-form errors, $\Omega_u^{(b)}$, where the superscript $(b)$ is the bootstrap index that runs from 1 to 500. The Choleski decomposition of $\Omega_u^{(b)}$ is denoted as $\Omega_{tr}^{(b)} = chol(\Omega_u^{(b)})$.

2. Aim at obtaining a candidate matrix $S^{(b,i)}$ whose first 2 columns satisfy the magnitude restrictions. The superscript $(i)$ would index the draw that runs from 1 to 100.

   (2a) Draw a 2x2 matrix from a standard normal distribution and obtain its QR decomposition, where the orthogonal matrix in the decomposition is denoted as $Q^{(b,i)}$, i.e. $Q^{(b,i)}Q^{(b,i)'} = I$.

   (2b) Construct the block diagonal matrix $Q_{diag}^{(b,i)} = diag(Q^{(b,i)}, I)$, with a size that corresponds to the size of $\Omega_u^{(b)}$.

   (2c) Rotate $Q_{diag}^{(b,i)}$ by a small rotation matrix constructed as $(I - H^{(b,i)})(I + H^{(b,i)})^{-1}$, where $H^{(b,i)}$ is hemisymmetric, i.e. $H^{(b,i)} = -H^{(b,i)'}$, and its elements are drawn from a random normal distribution. The new matrix $\tilde{Q}^{(b,i)} = Q_{diag}^{(b,i)}(I - H^{(b,i)})(I + H^{(b,i)})^{-1}$ will be orthogonal.

   (2d) Check whether the matrix $S^{(b,i)} = \Omega_{tr}^{(b)}\tilde{Q}^{(b,i)}$ satisfies the magnitude restrictions on its first 2 columns (i.e. the (1,1) element of $S^{(b,i)}$ has the highest absolute value among all elements in the first column; the (2,2) element of $S^{(b,i)}$ has the highest absolute value among all elements in the second column). If it does, we keep this $(i)$ draw for $S^{(b,i)}$. If not, we go back to step (2a). We repeat this process for 100 times to obtain a sufficient number of successful draws.

4. Repeat step 1 and 2 for 500 times; compute the 68% confidence bands of the IFRs after considering all successful candidate matrices $S^{(b,i)}$ from step 2d).

## Appendix 4: IRFs to uncertainty shocks identified through magnitude restrictions

**Figure A4.1**: IRFs to Italian uncertainty shocks

**Panel A**

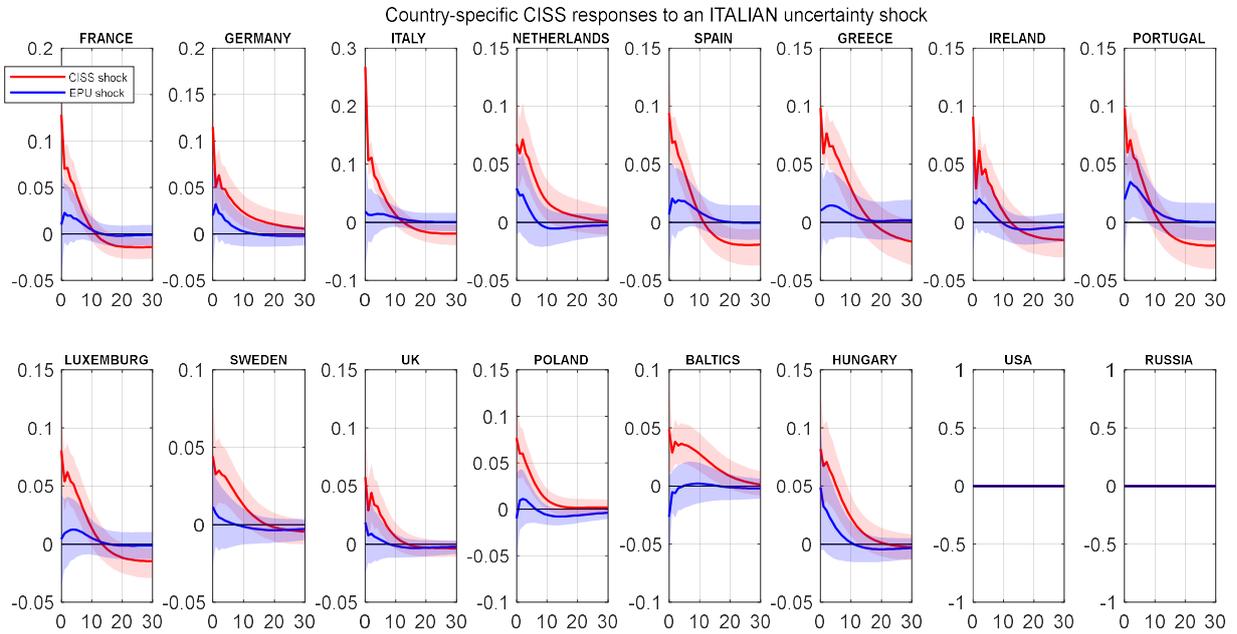

**Panel B**

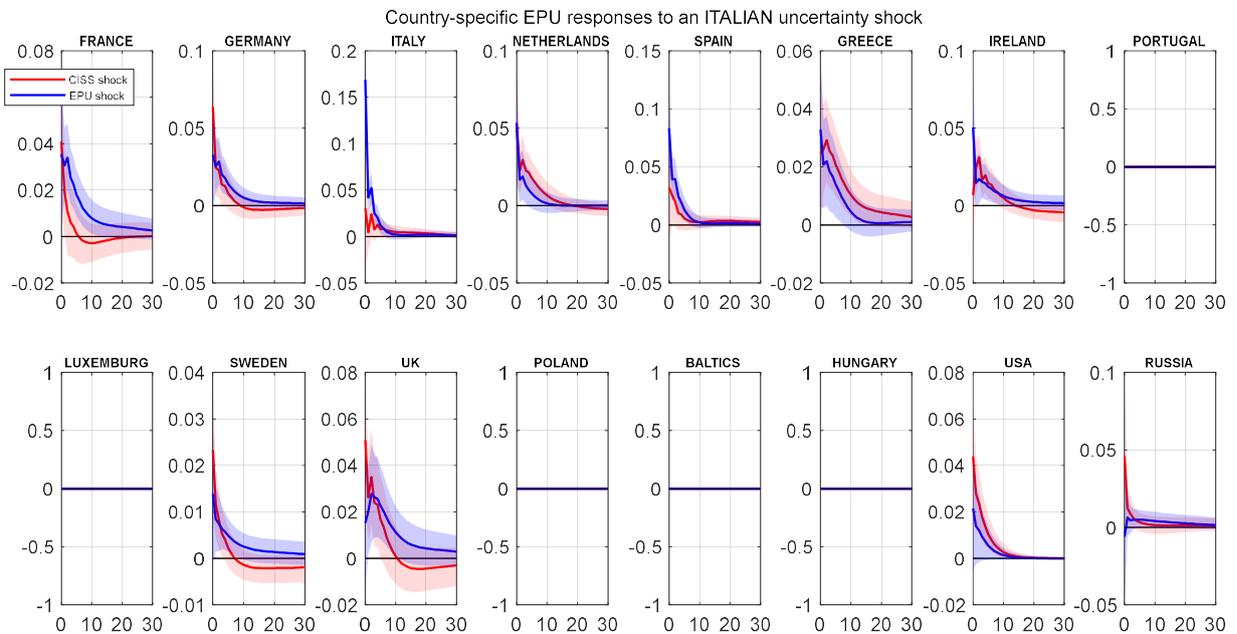

**Panel C**

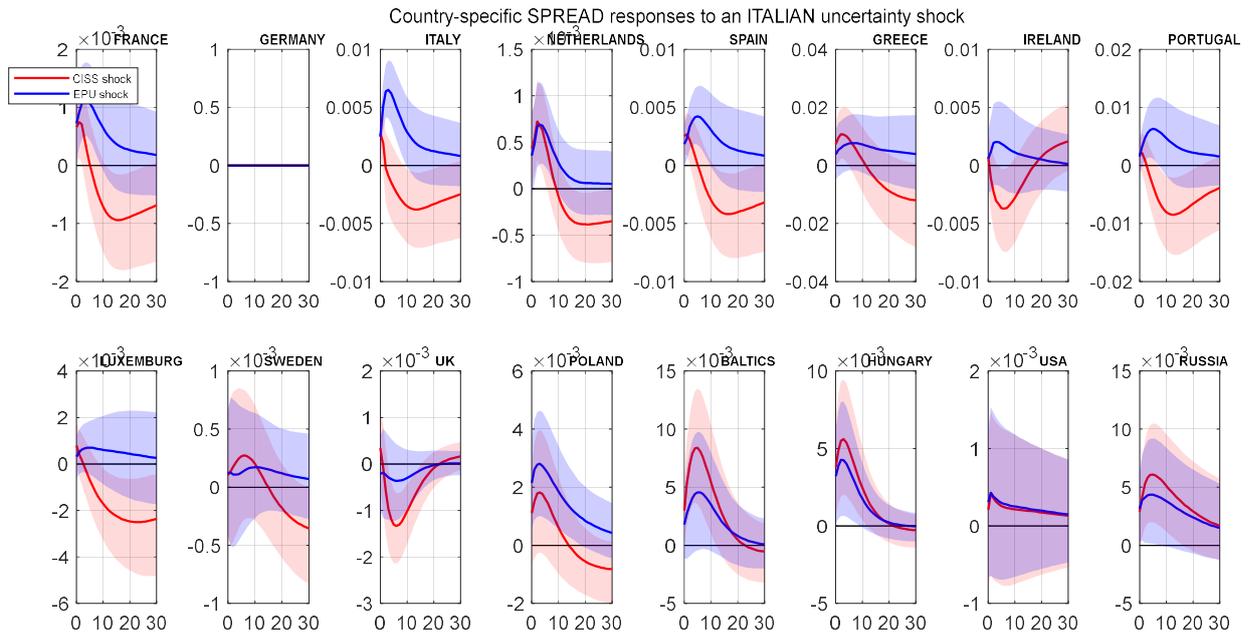

**Panel D**

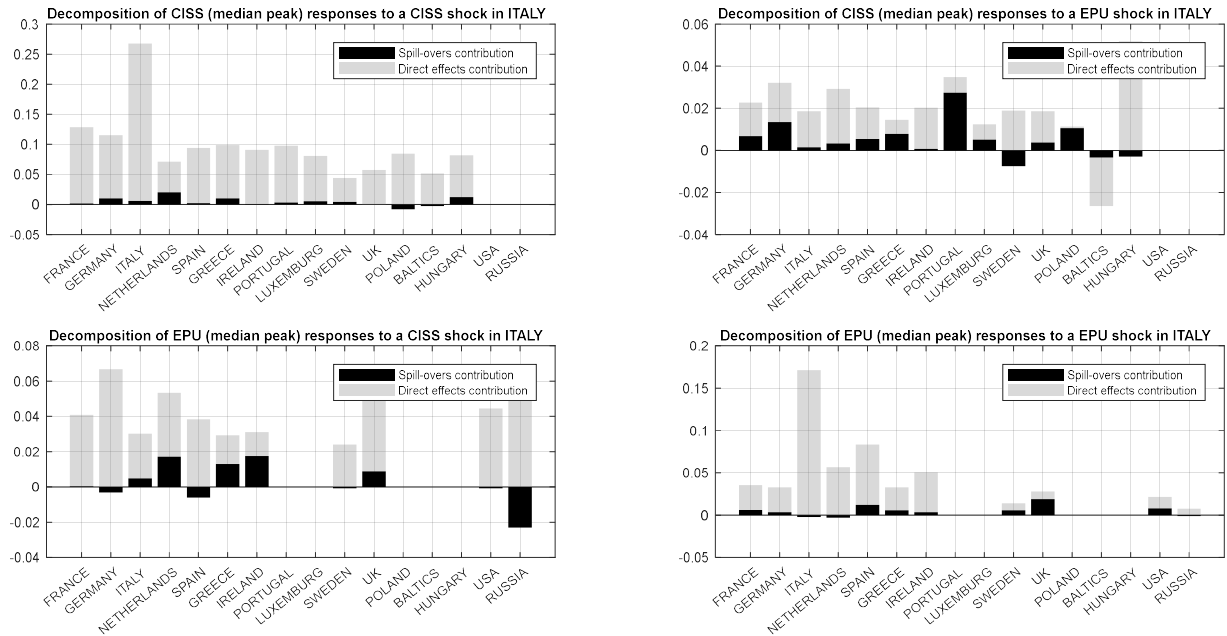

Note: Panels A, B and C plot the IRFs to both EPU and CISS uncertainty shocks (of 1 standard deviation). The 68% confidence bands are constructed from 500 bootstrapped replications of the GVAR, each with 100 maximum draws for the orthogonal matrix. Panel D displays the contribution from both direct effects and spill-overs to the peak response (calculated over the first 6 months) in the uncertainty variable indicated in the title of each subplot. Direct effects are calculated by turning off the country-specific foreign variables in the GVAR model, i.e. by setting the $C_{i,j}$ coefficients to zero in equation (1); spill-overs therefore are the difference between direct effects and the total effects, which are both approximated by the median bootstrapped IRFs.

**Figure A4.2**: IRFs to Spanish uncertainty shocks

**Panel A**

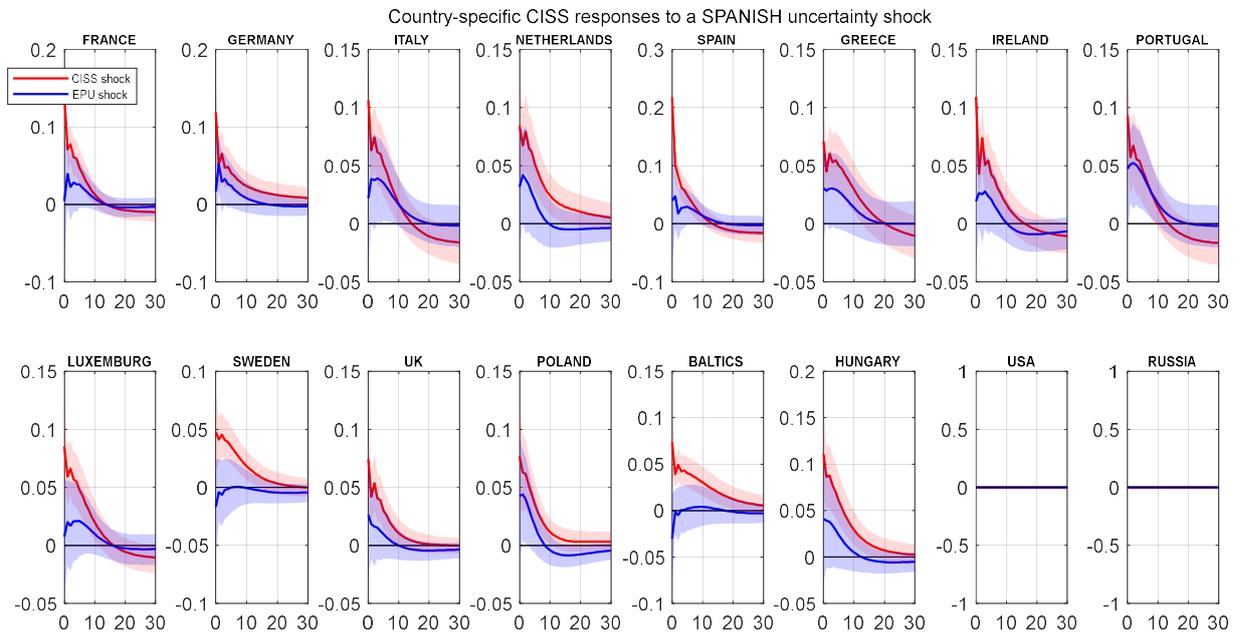

**Panel B**

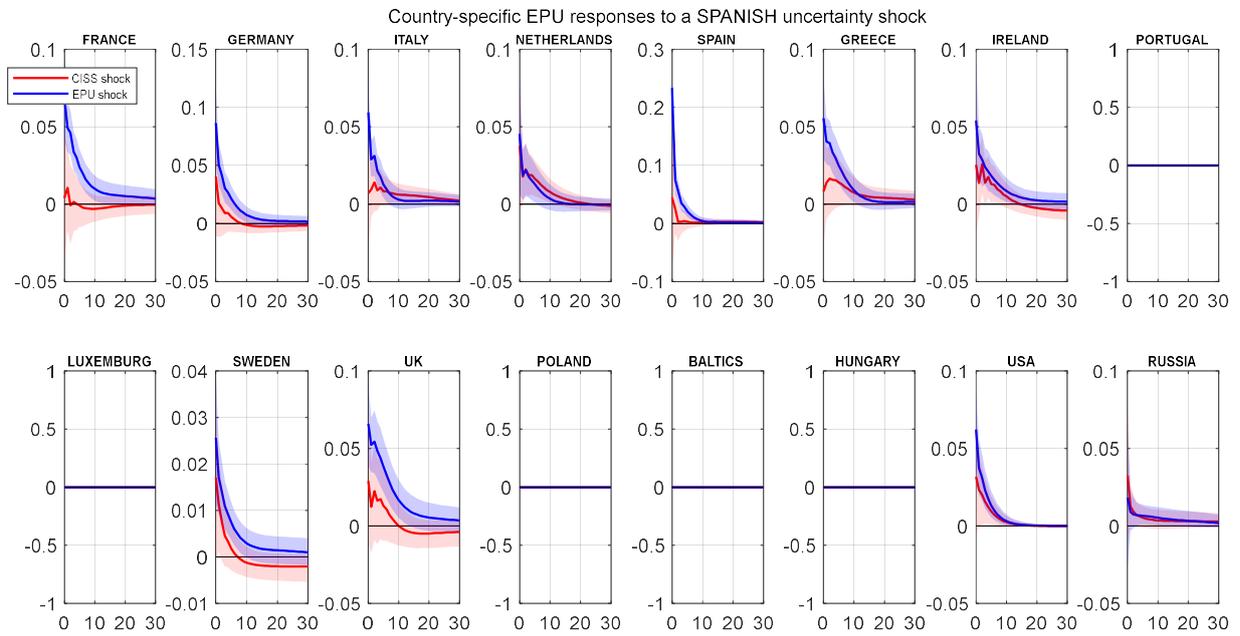

**Panel C**

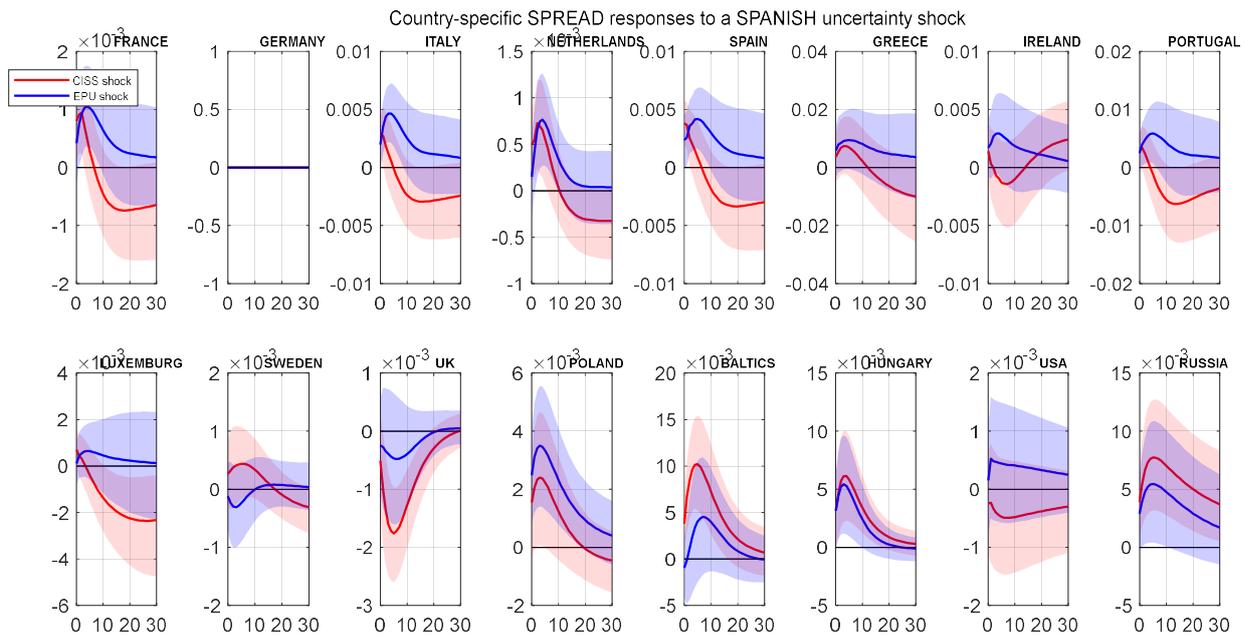

**Panel D**

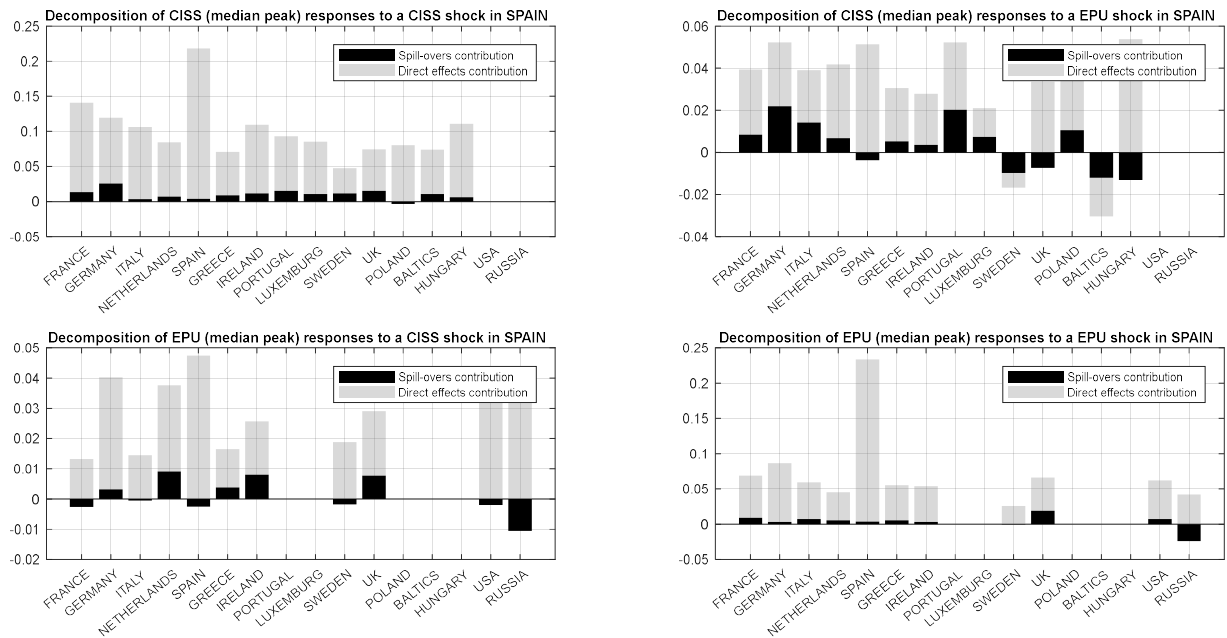

Note: Panels A, B and C plot the IRFs to both EPU and CISS uncertainty shocks (of 1 standard deviation). The 68% confidence bands are constructed from 500 bootstrapped replications of the GVAR, each with 100 maximum draws for the orthogonal matrix. Panel D displays the contribution from both direct effects and spill-overs to the peak response (calculated over the first 6 months) in the uncertainty variable indicated in the title of each subplot. Direct effects are calculated by turning off the country-specific foreign variables in the GVAR model, i.e. by setting the $C_{i,j}$ coefficients to zero in equation (1); spill-overs therefore are the difference between direct effects and the total effects, which are both approximated by the median bootstrapped IRFs.

**Figure A4.3**: IRFs to Greek uncertainty shocks

**Panel A**

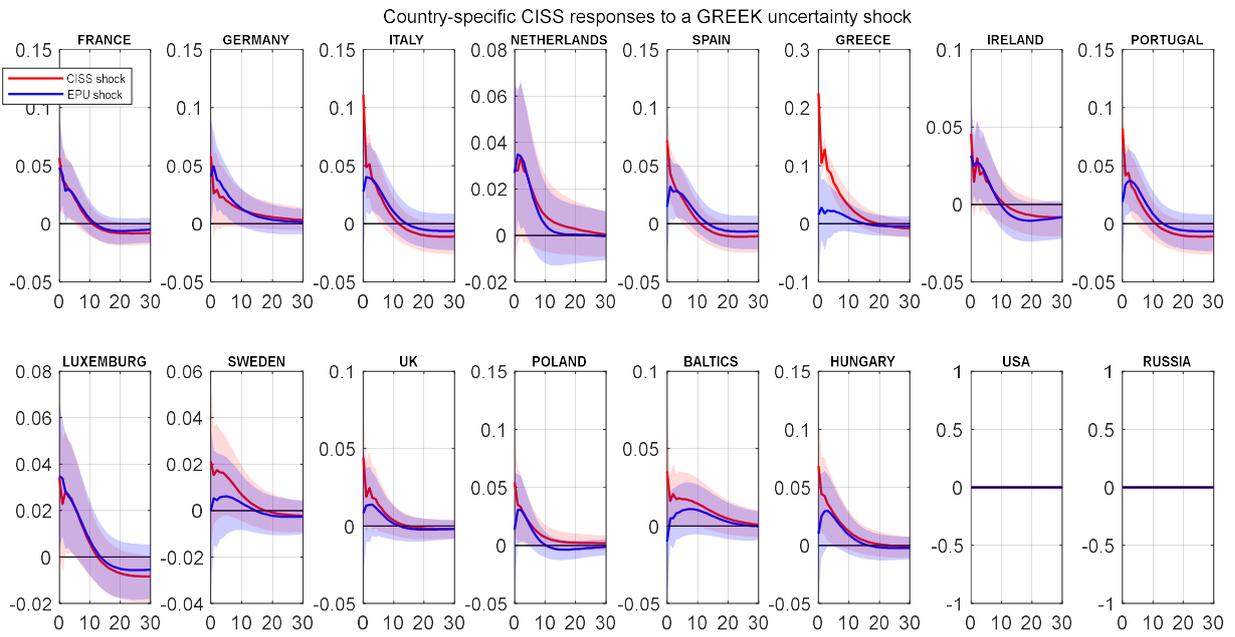

**Panel B**

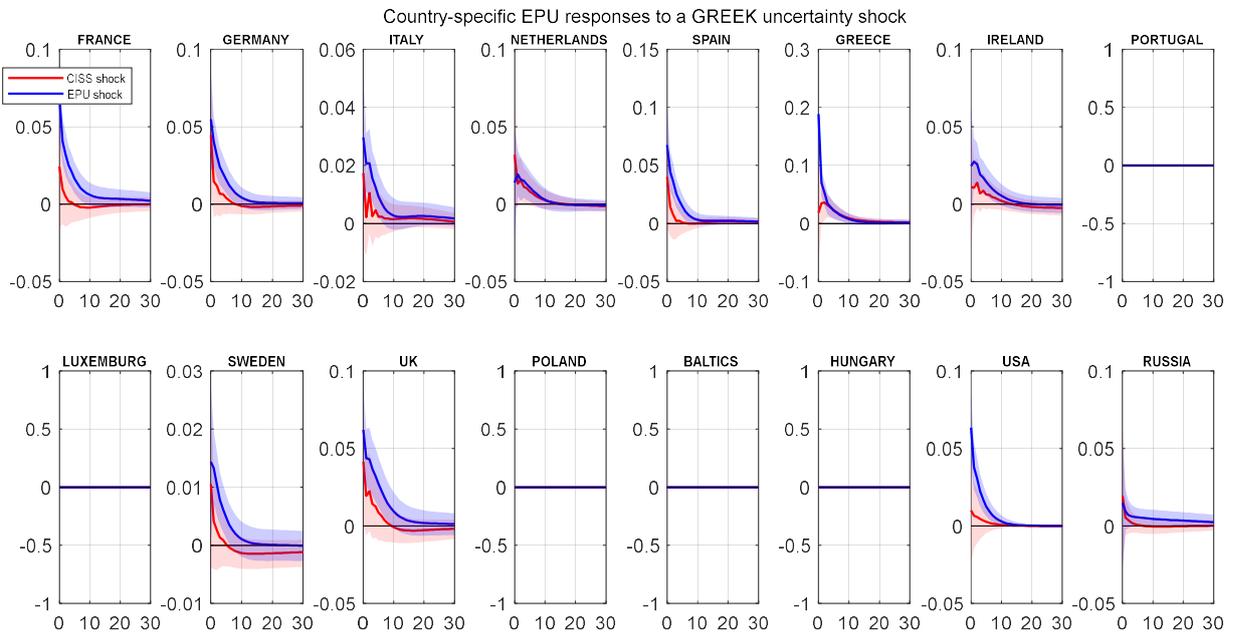

**Panel C**

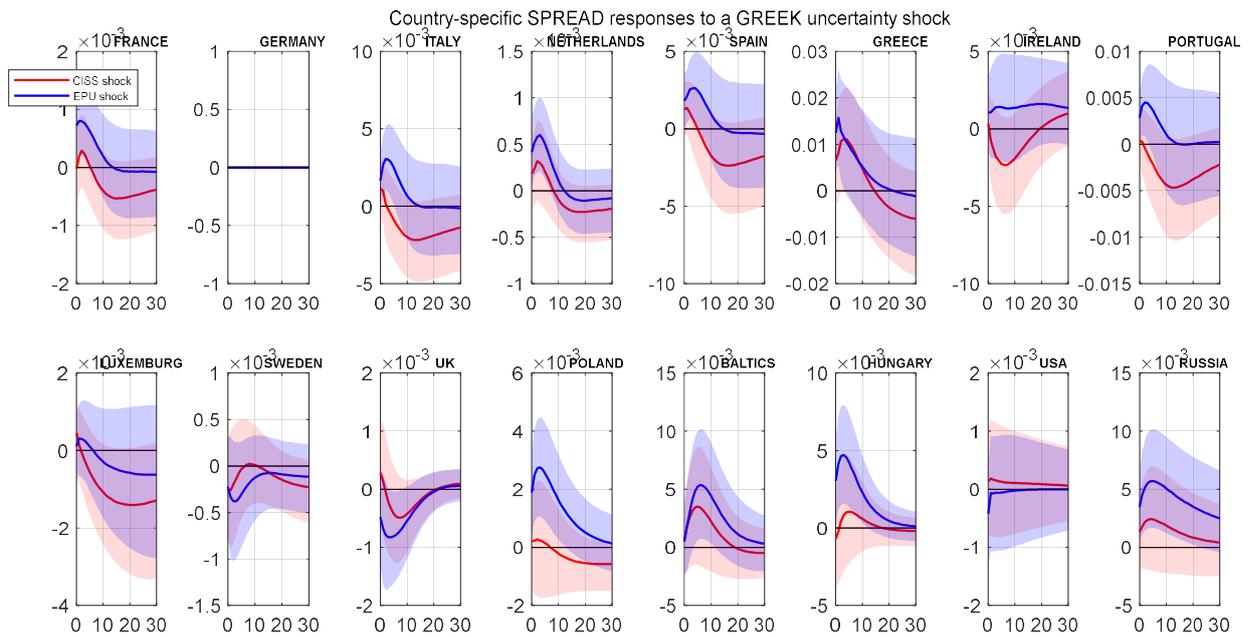

**Panel D**

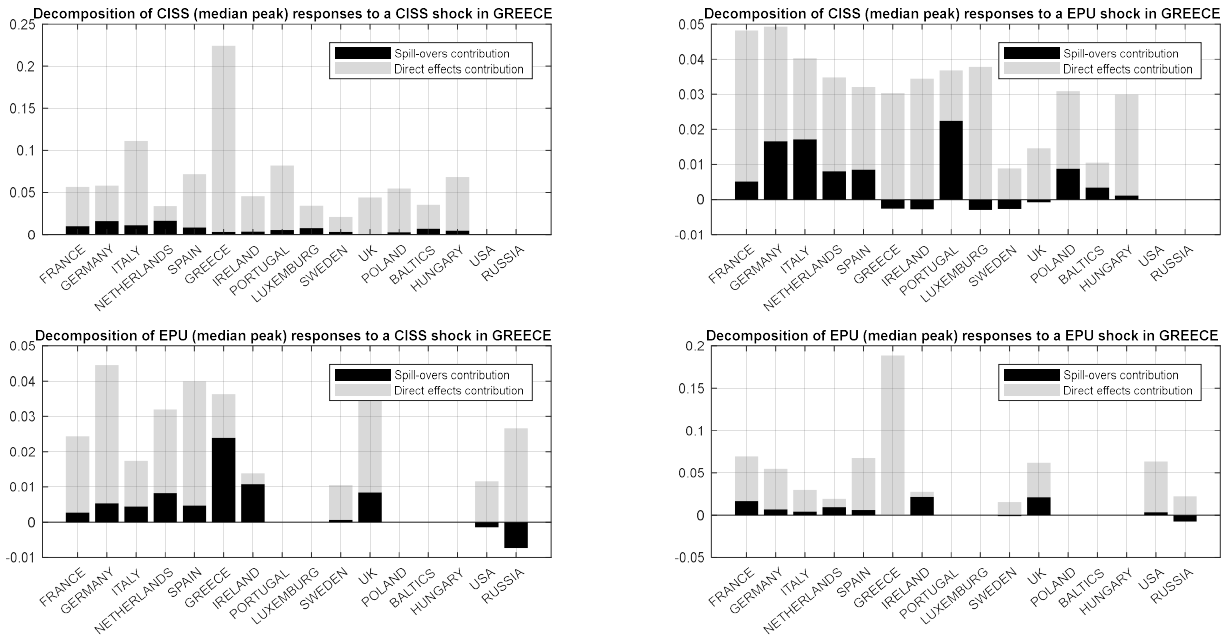

Note: Panels A, B and C plot the IRFs to both EPU and CISS uncertainty shocks (of 1 standard deviation). The 68% confidence bands are constructed from 500 bootstrapped replications of the GVAR, each with 100 maximum draws for the orthogonal matrix. Panel D displays the contribution from both direct effects and spill-overs to the peak response (calculated over the first 6 months) in the uncertainty variable indicated in the title of each subplot. Direct effects are calculated by turning off the country-specific foreign variables in the GVAR model, i.e. by setting the $C_{i,j}$ coefficients to zero in equation (1); spill-overs therefore are the difference between direct effects and the total effects, which are both approximated by the median bootstrapped IRFs.

**Figure A4.4**: IRFs to Irish uncertainty shocks

**Panel A**

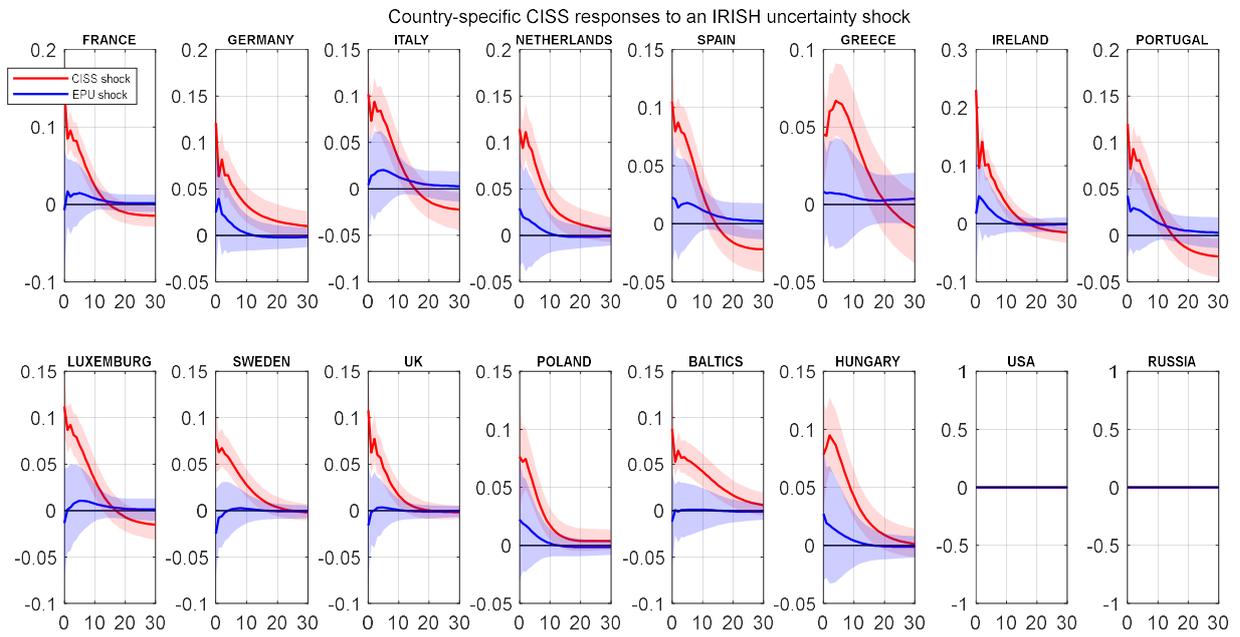

**Panel B**

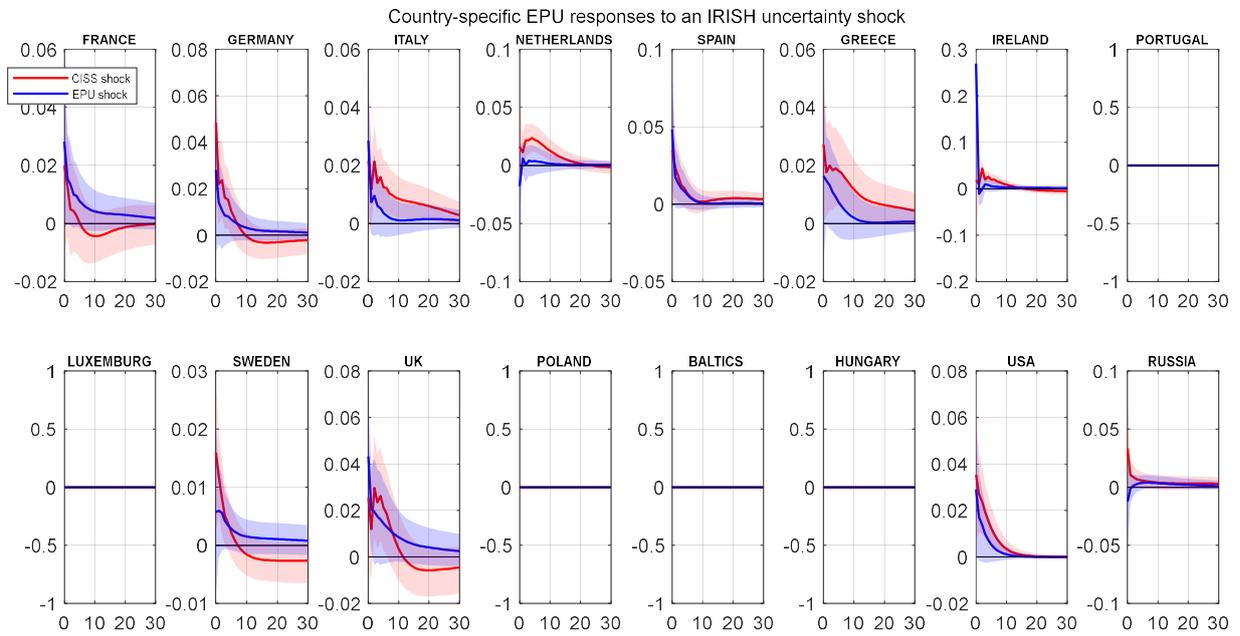

**Panel C**

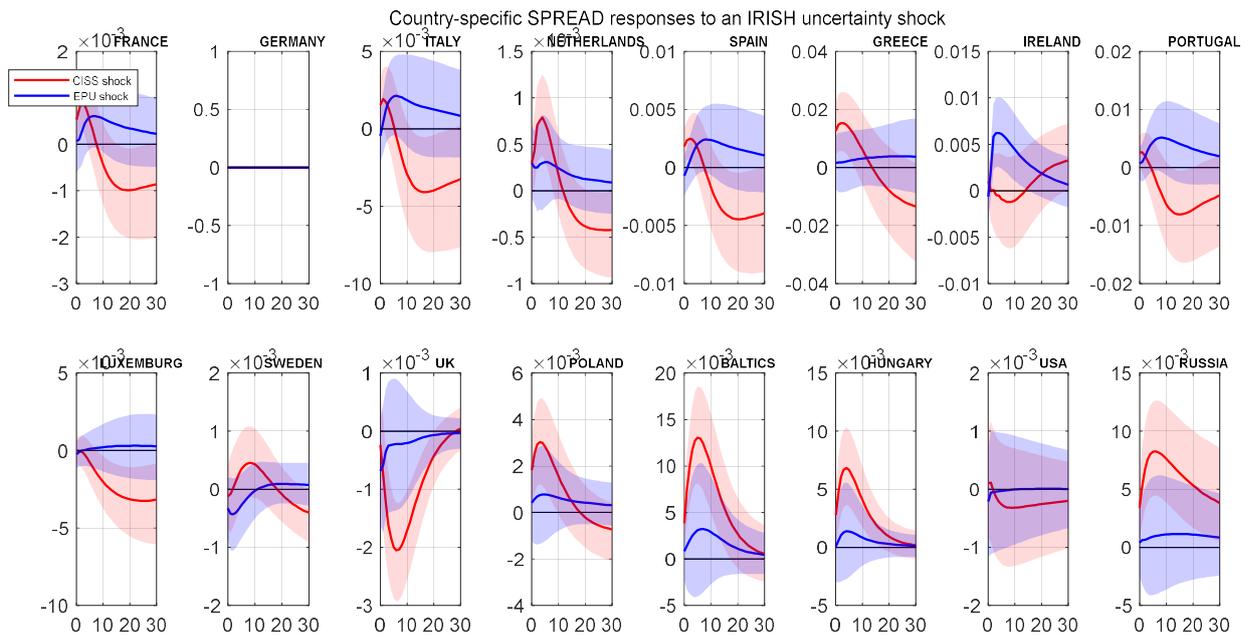

**Panel D**

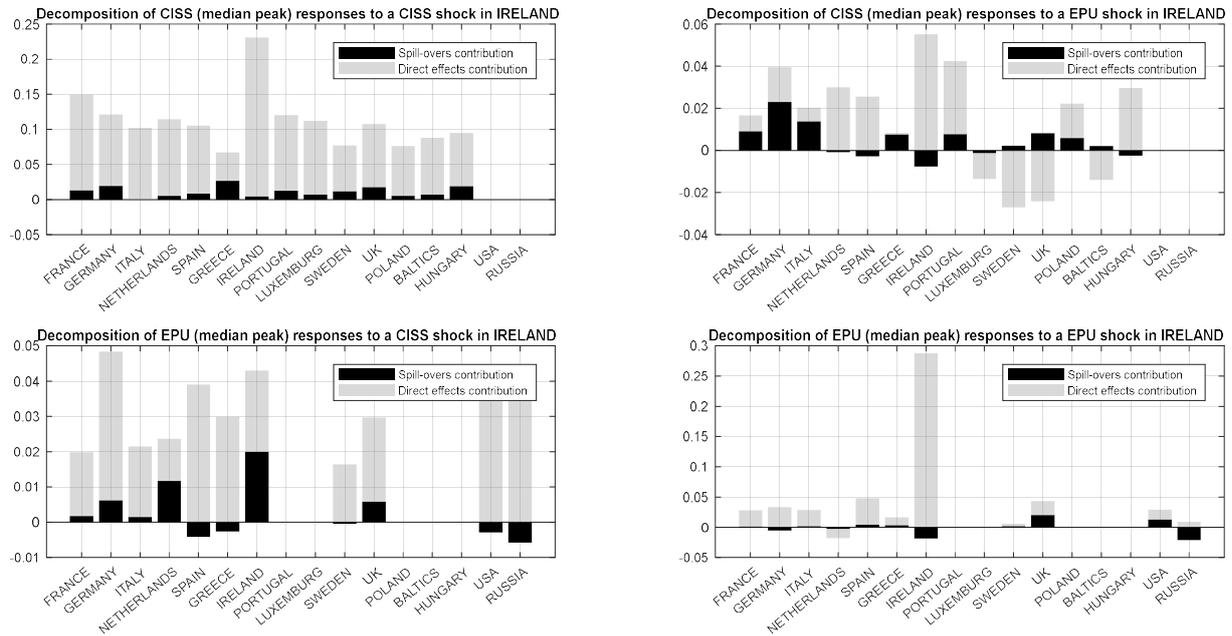

Note: Panels A, B and C plot the IRFs to both EPU and CISS uncertainty shocks (of 1 standard deviation). The 68% confidence bands are constructed from 500 bootstrapped replications of the GVAR, each with 100 maximum draws for the orthogonal matrix. Panel D displays the contribution from both direct effects and spill-overs to the peak response (calculated over the first 6 months) in the uncertainty variable indicated in the title of each subplot. Direct effects are calculated by turning off the country-specific foreign variables in the GVAR model, i.e. by setting the $C_{i,j}$ coefficients to zero in equation (1); spill-overs therefore are the difference between direct effects and the total effects, which are both approximated by the median bootstrapped IRFs.

**Figure A4.5**: IRFs to French uncertainty shocks

**Panel A**

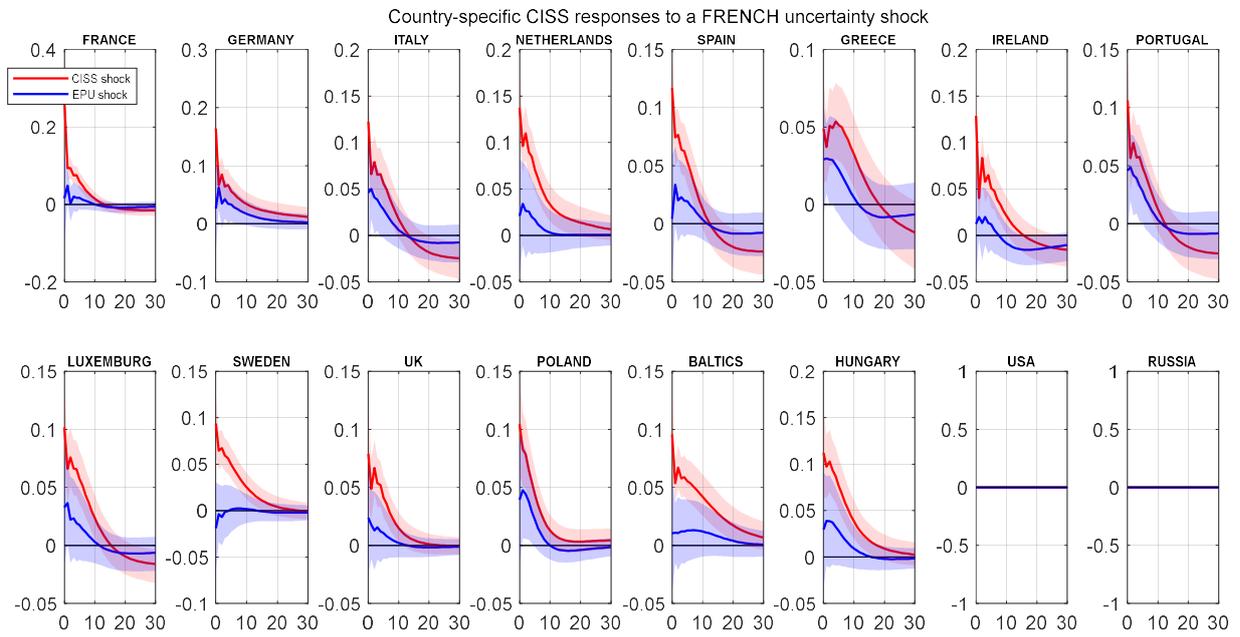

**Panel B**

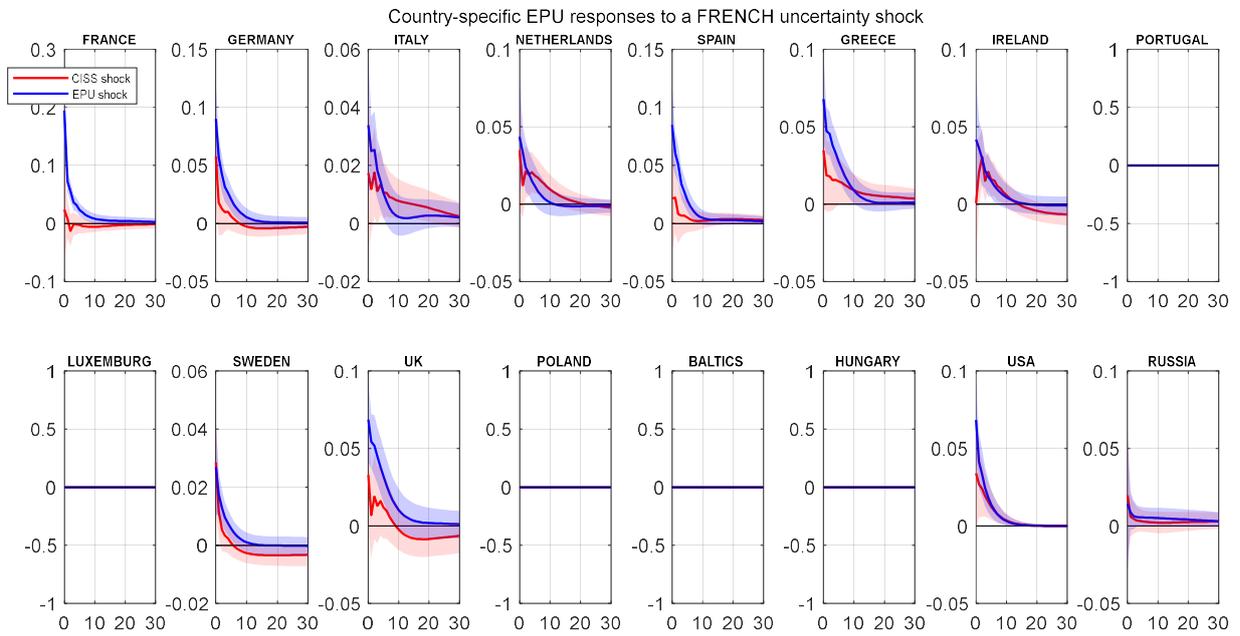

**Panel C**

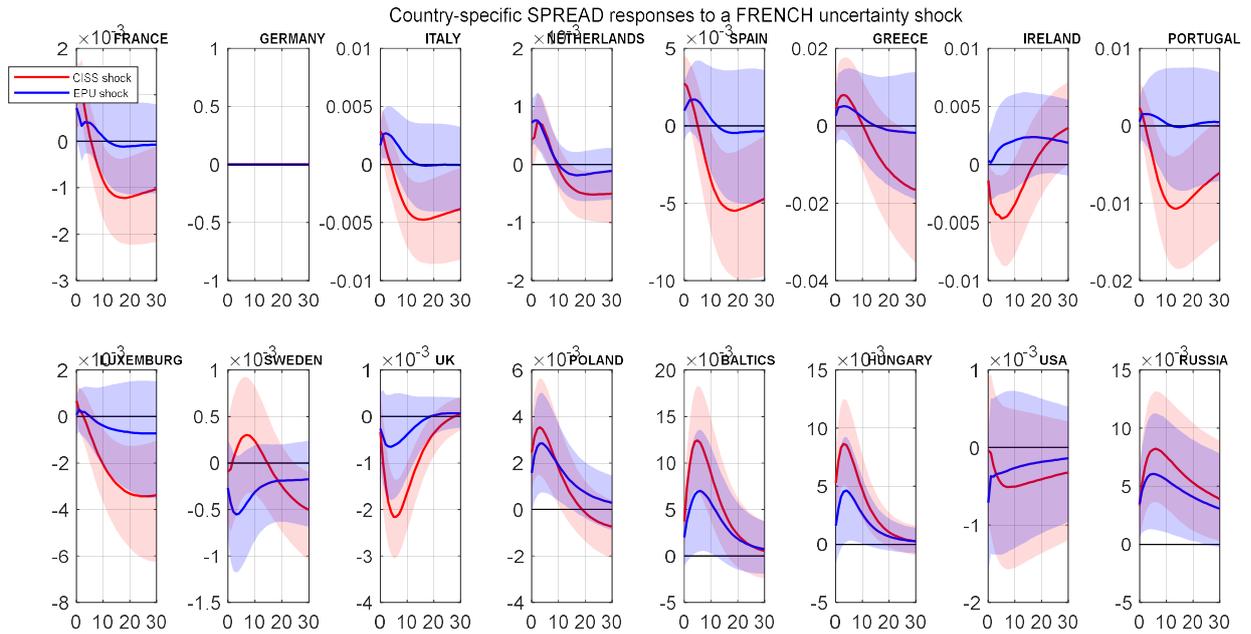

**Panel D**

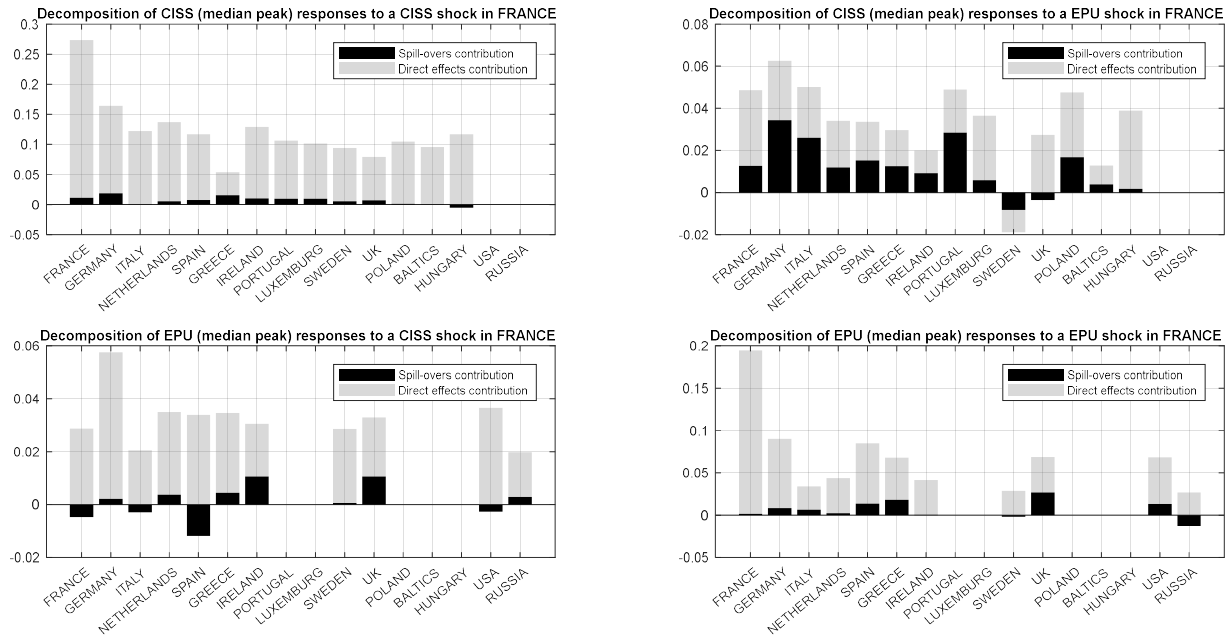

Note: Panels A, B and C plot the IRFs to both EPU and CISS uncertainty shocks (of 1 standard deviation). The 68% confidence bands are constructed from 500 bootstrapped replications of the GVAR, each with 100 maximum draws for the orthogonal matrix. Panel D displays the contribution from both direct effects and spill-overs to the peak response (calculated over the first 6 months) in the uncertainty variable indicated in the title of each subplot. Direct effects are calculated by turning off the country-specific foreign variables in the GVAR model, i.e. by setting the $C_{i,j}$ coefficients to zero in equation (1); spill-overs therefore are the difference between direct effects and the total effects, which are both approximated by the median bootstrapped IRFs.

## Appendix 5

**Figure A5**: IFRs for ECB monetary policy proxies to uncertainty shocks

**Panel A**

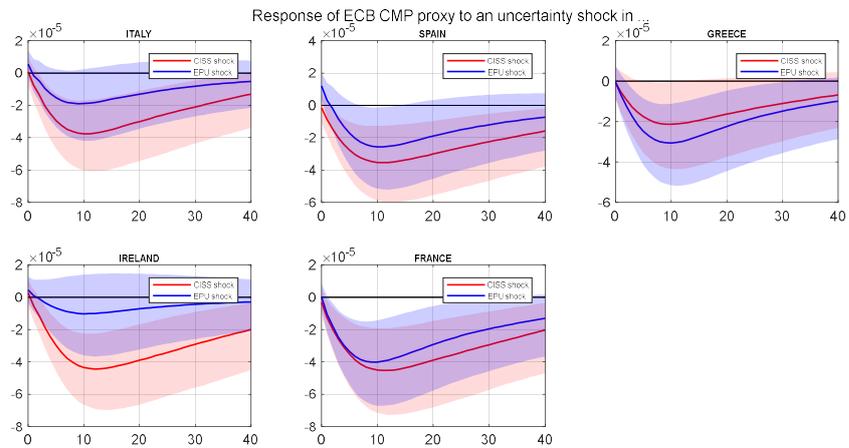

**Panel B**

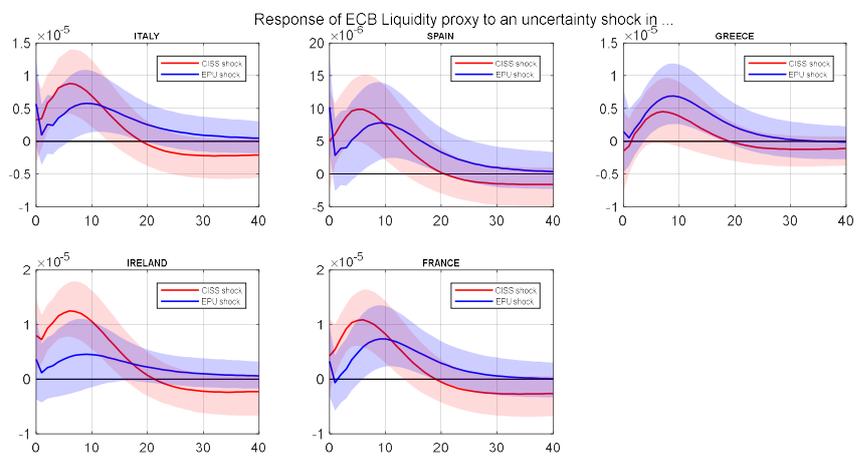

**Panel C**

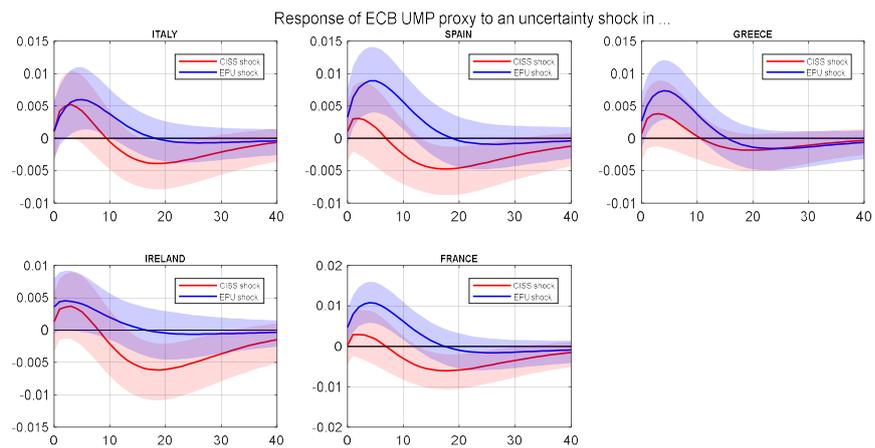

Note: The title of the plots in each panel displays the origin country of the uncertainty shock. The legend displays the corresponding uncertainty shock that is being simulated. The 68% confidence bands are constructed from 500 bootstrapped replications of the GVAR, each with 100 maximum draws for the orthogonal matrix (see algorithm in Appendix 3).

**Appendix 6**

**Figure A6**: Histograms of the uncertainty shocks in countries where identification is performed

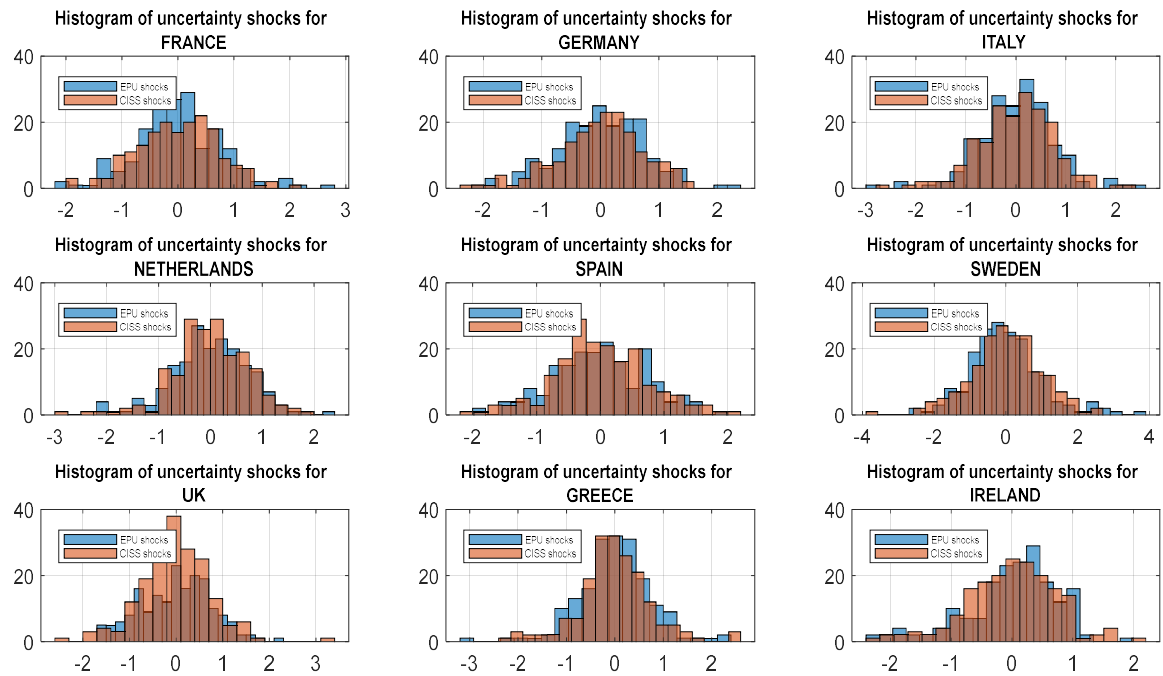